%% file: main.tex
\pdfoutput=1
\documentclass[12pt,a4paper]{article}

\usepackage{ifthen} 
\usepackage{comment}
\usepackage{multirow}
\newboolean{pdflatex}
\setboolean{pdflatex}{true} 

\newboolean{articletitles}
\setboolean{articletitles}{true} 

\newboolean{uprightparticles}
\setboolean{uprightparticles}{false} 

\def\paperauthors{LHCb collaboration} 
\def\paperasciititle{Observation and investigation of the Tccbar1(4430)+ structure in B+ -> psi(2S) KS0 pi+ decays} 
\def\papertitle{Observation and investigation of the $T_{c\bar{c}1}(4430)^{+}$ structure in $B^{+} \to \psi(2S) K_{\text{S}}^{0} \pi^{+}$ decays} 
\def\paperkeywords{{High Energy Physics}, {LHCb}} 
\def\papercopyright{\the\year\ CERN for the benefit of the LHCb collaboration} 
\def\paperlicence{CC BY 4.0 licence}
\def\paperlicenceurl{https://creativecommons.org/licenses/by/4.0/}

\input{preamble}
\usepackage{longtable} 

\begin{document}

\renewcommand{\thefootnote}{\fnsymbol{footnote}}
\setcounter{footnote}{1}

\input{title-LHCb-PAPER}


\renewcommand{\thefootnote}{\arabic{footnote}}
\setcounter{footnote}{0}

\cleardoublepage


\pagestyle{plain} 
\setcounter{page}{1}
\pagenumbering{arabic}



\input{body}

\input{acknowledgements}



\input{supplementary-app}

\addcontentsline{toc}{section}{References}
\bibliographystyle{LHCb}
\bibliography{main,standard,LHCb-PAPER,LHCb-CONF,LHCb-DP,LHCb-TDR}

\newpage
\input{Authorship_LHCb-PAPER-2025-039}

\end{document}

%% file: preamble.tex
\usepackage[top=1in, bottom=1.25in, left=1in, right=1in]{geometry}

\usepackage{microtype}
\usepackage{lineno}  
\usepackage{xspace} 
\usepackage{caption} 

\usepackage{graphicx}  
\usepackage{color}
\usepackage{colortbl}
\graphicspath{{./figs/}} 

\usepackage{amsmath} 
\usepackage{amssymb}
\usepackage{amsfonts}
\usepackage{upgreek} 

\newcommand*\patchAmsMathEnvironmentForLineno[1]{%
\expandafter\let\csname old#1\expandafter\endcsname\csname #1\endcsname
\expandafter\let\csname oldend#1\expandafter\endcsname\csname
end#1\endcsname
 \renewenvironment{#1}%
   {\linenomath\csname old#1\endcsname}%
   {\csname oldend#1\endcsname\endlinenomath}%
}
\newcommand*\patchBothAmsMathEnvironmentsForLineno[1]{%
  \patchAmsMathEnvironmentForLineno{#1}%
  \patchAmsMathEnvironmentForLineno{#1*}%
}
\AtBeginDocument{%
\patchBothAmsMathEnvironmentsForLineno{equation}%
\patchBothAmsMathEnvironmentsForLineno{align}%
\patchBothAmsMathEnvironmentsForLineno{flalign}%
\patchBothAmsMathEnvironmentsForLineno{alignat}%
\patchBothAmsMathEnvironmentsForLineno{gather}%
\patchBothAmsMathEnvironmentsForLineno{multline}%
\patchBothAmsMathEnvironmentsForLineno{eqnarray}%
}

\usepackage{hyperxmp}

\usepackage[pdftex,
            pdfauthor={\paperauthors},
            pdftitle={\paperasciititle},
            pdfkeywords={\paperkeywords},
            pdfcopyright={Copyright (C) \papercopyright},
            pdflicenseurl={\paperlicenceurl}]{hyperref}

\usepackage[colorinlistoftodos,textsize=scriptsize]{todonotes}

\usepackage[bottom,flushmargin,hang,multiple]{footmisc}

\usepackage[all]{hypcap} 

\input{lhcb-symbols-def} 

\usepackage{cite} 
\usepackage{mciteplus}

%% file: lhcb-symbols-def.tex
\usepackage{xspace} 
\usepackage{upgreek}

\def\lhcb   {\mbox{LHCb}\xspace}

\def\MagUp {\mbox{\em Mag\kern -0.05em Up}\xspace}

\ifthenelse{\boolean{uprightparticles}}%
{

 \def\Pmu         {\ensuremath{\upmu}\xspace}

 \def\Ppi         {\ensuremath{\uppi}\xspace}

 \def\Ppsi        {\ensuremath{\uppsi}\xspace}

 \def\PDelta      {\ensuremath{\Delta}\xspace}                 
 \def\PXi         {\ensuremath{\Xi}\xspace}                 
 \def\PLambda     {\ensuremath{\Lambda}\xspace}                 
 \def\PSigma      {\ensuremath{\Sigma}\xspace}                 
 \def\POmega      {\ensuremath{\Omega}\xspace}                 
 \def\PUpsilon    {\ensuremath{\Upsilon}\xspace}
 \let\oldPi\Pi
 \def\PPi         {\ensuremath{\oldPi}\xspace}

 \def\PB      {\ensuremath{\mathrm{B}}\xspace}                 
                  
 \def\PD      {\ensuremath{\mathrm{D}}\xspace}

 \def\PK      {\ensuremath{\mathrm{K}}\xspace}

 \def\Pi      {\ensuremath{\mathrm{i}}\xspace}

 \def\Ps      {\ensuremath{\mathrm{s}}\xspace}

 \def\thebaroffset{0.0em}
}
{

 \def\Pmu         {\ensuremath{\mu}\xspace}

 \def\Ppi         {\ensuremath{\pi}\xspace}

 \def\Ppsi        {\ensuremath{\psi}\xspace}                 
                  
 \mathchardef\PDelta="7101
 \mathchardef\PXi="7104
 \mathchardef\PLambda="7103
 \mathchardef\PSigma="7106
 \mathchardef\POmega="710A
 \mathchardef\PUpsilon="7107
 \mathchardef\PPi="7105
                  
 \def\PB      {\ensuremath{B}\xspace}                 
                  
 \def\PD      {\ensuremath{D}\xspace}

 \def\PK      {\ensuremath{K}\xspace}

 \def\Pi      {\ensuremath{i}\xspace}

 \def\Ps      {\ensuremath{s}\xspace}

 \def\thebaroffset{0.18em}
}
\newcommand{\offsetoverline}[2][\thebaroffset]{\kern #1\overline{\kern -#1 #2}}%

\makeatletter
\ifcase \@ptsize \relax
  \newcommand{\miniscule}{\@setfontsize\miniscule{4}{5}}
\or
  \newcommand{\miniscule}{\@setfontsize\miniscule{5}{6}}
\or
  \newcommand{\miniscule}{\@setfontsize\miniscule{5}{6}}
\fi
\makeatother

\DeclareRobustCommand{\optbar}[1]{\shortstack{{\miniscule (\rule[.5ex]{1.25em}{.18mm})}
  \\ [-.7ex] $#1$}}

\def\mup        {{\ensuremath{\Pmu^+}}\xspace}
\def\mun        {{\ensuremath{\Pmu^-}}\xspace} 

\def\squark    {{\ensuremath{\Ps}}\xspace}

\def\pion   {{\ensuremath{\Ppi}}\xspace}

\def\pip    {{\ensuremath{\pion^+}}\xspace}
\def\pim    {{\ensuremath{\pion^-}}\xspace}

\def\kaon    {{\ensuremath{\PK}}\xspace}
\def\Kbar    {{\ensuremath{\offsetoverline{\PK}}}\xspace}

\def\KorKbar {\kern \thebaroffset\optbar{\kern -\thebaroffset \PK}{}\xspace}

\def\Km      {{\ensuremath{\kaon^-}}\xspace}

\def\KS      {{\ensuremath{\kaon^0_{\mathrm{S}}}}\xspace}

\def\Kstar   {{\ensuremath{\kaon^*}}\xspace}
\def\Kstarb  {{\ensuremath{\Kbar{}^*}}\xspace}
\def\Kstarp  {{\ensuremath{\kaon^{*+}}}\xspace}

\def\Dbar    {{\ensuremath{\offsetoverline{\PD}}}\xspace}
\def\D       {{\ensuremath{\PD}}\xspace}

\def\DorDbar {\kern \thebaroffset\optbar{\kern -\thebaroffset \PD}\xspace}

\def\Dp      {{\ensuremath{\D^+}}\xspace}
\def\Dm      {{\ensuremath{\D^-}}\xspace}

\def\DpDm    {\ensuremath{\Dp {\kern -0.16em \Dm}}\xspace}

\def\B       {{\ensuremath{\PB}}\xspace}
\def\Bbar    {{\ensuremath{\offsetoverline{\PB}}}\xspace}

\def\BorBbar {\kern \thebaroffset\optbar{\kern -\thebaroffset \PB}\xspace}

\def\Bzb     {{\ensuremath{\Bbar{}^0}}\xspace}
\def\Bd      {{\ensuremath{\B^0}}\xspace}

\def\BdorBdbar {\kern \thebaroffset\optbar{\kern -\thebaroffset \Bd}\xspace}
\def\Bu      {{\ensuremath{\B^+}}\xspace}

\def\Bp      {{\ensuremath{\Bu}}\xspace}

\def\Bs      {{\ensuremath{\B^0_\squark}}\xspace}

\def\BsorBsbar {\kern \thebaroffset\optbar{\kern -\thebaroffset \Bs}\xspace}

\def\psitwos  {{\ensuremath{\Ppsi{(2S)}}}\xspace}

\def\Y#1S{\ensuremath{\PUpsilon{(#1S)}}\xspace}

\def\LorLbar     {\kern \thebaroffset\optbar{\kern -\thebaroffset \PLambda}\xspace}

\newcommand{\decay}[2]{\ensuremath{#1\!\to #2}\xspace} 

\def\to                 {\ensuremath{\rightarrow}\xspace}

\def\AT#1     {\ensuremath{A_{\mathrm{T}}^{#1}}\xspace}           

\def\C#1      {\ensuremath{\mathcal{C}_{#1}}\xspace}                       
\def\Cp#1     {\ensuremath{\mathcal{C}_{#1}^{'}}\xspace}                    
\def\Ceff#1   {\ensuremath{\mathcal{C}_{#1}^{\mathrm{(eff)}}}\xspace}        
\def\Cpeff#1  {\ensuremath{\mathcal{C}_{#1}^{'\mathrm{(eff)}}}\xspace}       
\def\Ope#1    {\ensuremath{\mathcal{O}_{#1}}\xspace}                       
\def\Opep#1   {\ensuremath{\mathcal{O}_{#1}^{'}}\xspace}                    

\newcommand{\aunit}[1]{\ensuremath{\text{\,#1}}}       

\newcommand{\tev}{\aunit{Te\kern -0.1em V}\xspace}
\newcommand{\gev}{\aunit{Ge\kern -0.1em V}\xspace}
\newcommand{\mev}{\aunit{Me\kern -0.1em V}\xspace}
\newcommand{\kev}{\aunit{ke\kern -0.1em V}\xspace}
\newcommand{\ev}{\aunit{e\kern -0.1em V}\xspace}
 
\newcommand{\mevc}{\ensuremath{\aunit{Me\kern -0.1em V\!/}c}\xspace}
\newcommand{\gevc}{\ensuremath{\aunit{Ge\kern -0.1em V\!/}c}\xspace}
\newcommand{\mevcc}{\ensuremath{\aunit{Me\kern -0.1em V\!/}c^2}\xspace}
\newcommand{\gevcc}{\ensuremath{\aunit{Ge\kern -0.1em V\!/}c^2}\xspace}

\def\fb   {\ensuremath{\aunit{fb}}\xspace}
\def\invfb   {\ensuremath{\fb^{-1}}\xspace}

\def\deriv {\ensuremath{\mathrm{d}}}

\def\gsim{{~\raise.15em\hbox{$>$}\kern-.85em
          \lower.35em\hbox{$\sim$}~}\xspace}
\def\lsim{{~\raise.15em\hbox{$<$}\kern-.85em
          \lower.35em\hbox{$\sim$}~}\xspace}

\newcommand{\Real}{\ensuremath{\mathcal{R}e}\xspace}
\newcommand{\Imag}{\ensuremath{\mathcal{I}m}\xspace}

\def\tell1  {TELL1\xspace}
\def\ukl1   {UKL1\xspace}

\def\myDecay {\ensuremath{\decay{\Bp}{\psitwos \KS \pip}}\xspace}
\def\myZcp {\ensuremath{T_{c\bar{c}1}(4430)^+}\xspace}

\def\myZc  {\ensuremath{T_{c\bar{c}}^+}\xspace}

\def\mydst  {\ensuremath{\Dbar{}^*_1(2600)^0}\xspace}
\def\mydd  {\ensuremath{\Dbar{}^*_1(2600)^0 D^+}\xspace}

\newcommand{\lhcborcid}[1]{\href{https://orcid.org/#1}{\hspace*{0.1em}\raisebox{-0.45ex}{\includegraphics[width=1em]{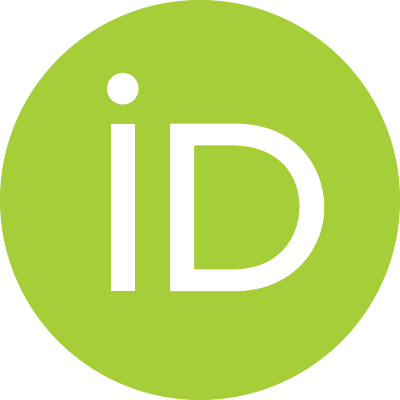}}}}


%% file: title-LHCb-PAPER.tex

\begin{titlepage}
\pagenumbering{roman}

\vspace*{-1.5cm}
\centerline{\large EUROPEAN ORGANIZATION FOR NUCLEAR RESEARCH (CERN)}
\vspace*{1.5cm}
\noindent
\begin{tabular*}{\linewidth}{lc@{\extracolsep{\fill}}r@{\extracolsep{0pt}}}
\ifthenelse{\boolean{pdflatex}}
{\vspace*{-1.5cm}\mbox{\!\!\!\includegraphics[width=.14\textwidth]{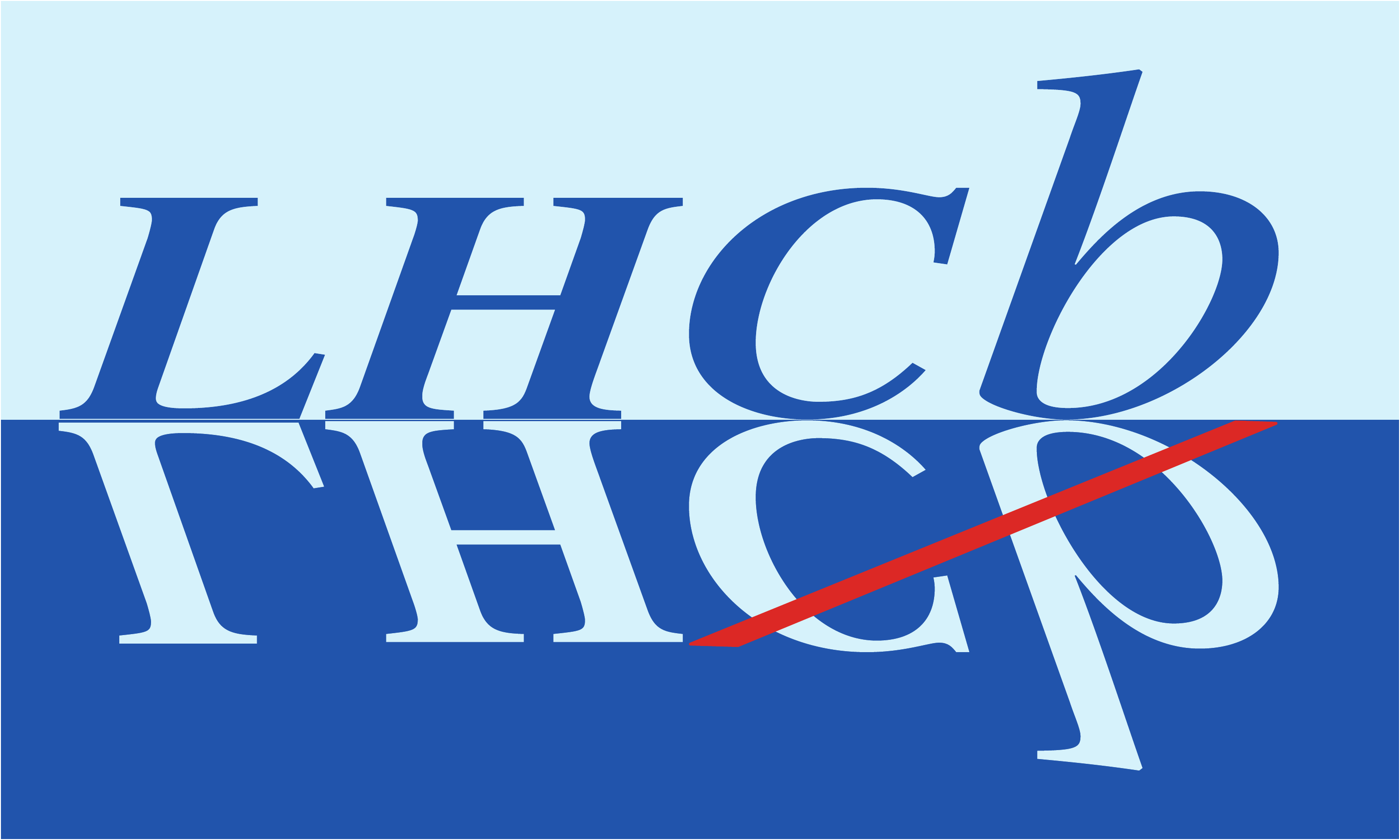}} & &}%
{\vspace*{-1.2cm}\mbox{\!\!\!\includegraphics[width=.12\textwidth]{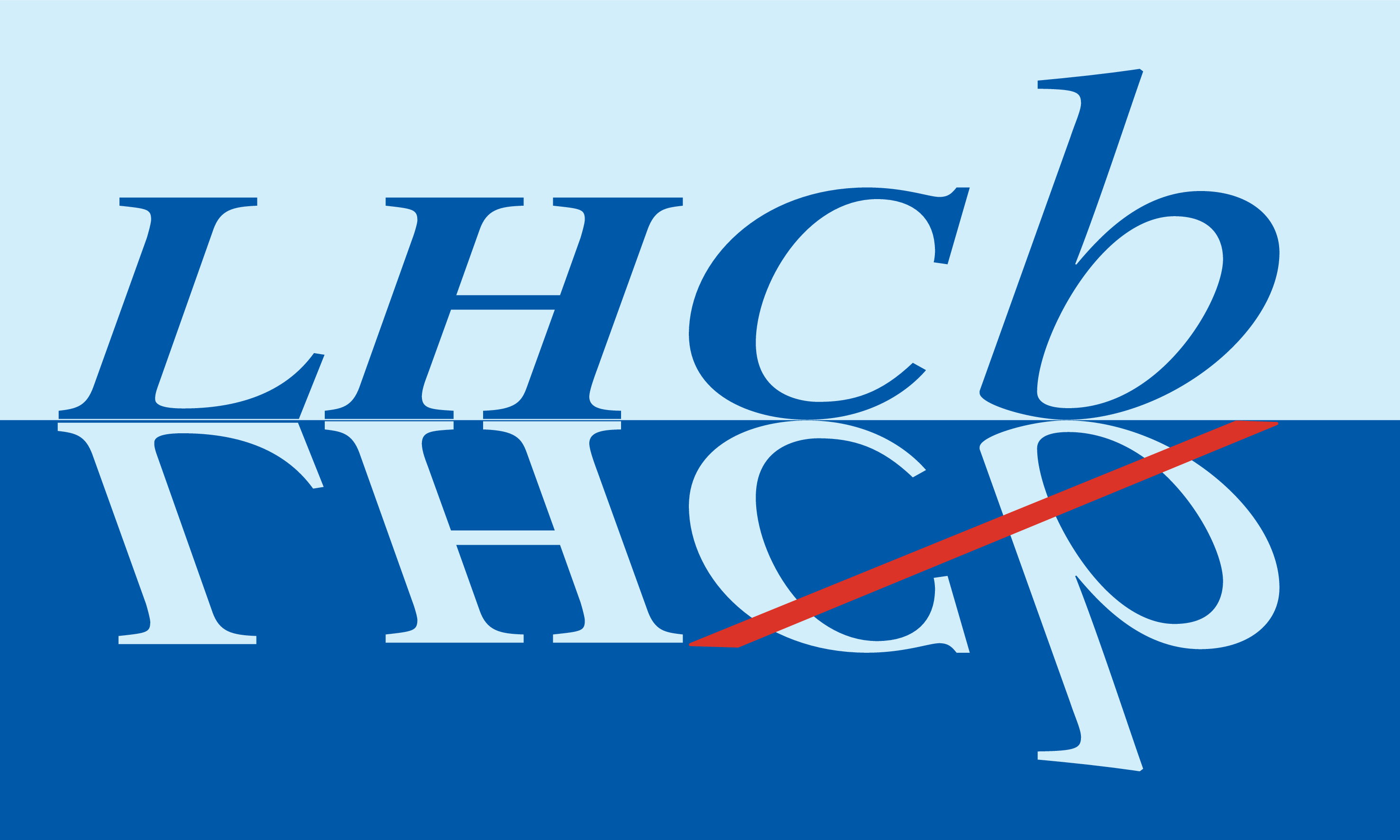}} & &}%
\\
 & & CERN-EP-2025-234 \\  
 & & LHCb-PAPER-2025-039 \\  
 & & April 7, 2026 \\ 
 & & \\
\end{tabular*}

\vspace*{2.0cm}

{\normalfont\bfseries\boldmath\huge
\begin{center}
  \papertitle 
\end{center}
}

\vspace*{2.0cm}

\begin{center}
\paperauthors\footnote{Authors are listed at the end of this paper.}
\end{center}

\vspace{\fill}

\begin{abstract}
  \noindent

The first four-dimensional amplitude analysis of the $B^{+} \to \psi(2S) K_{\text{S}}^{0} \pi^{+}$ decay is performed with proton-proton collision data collected by the LHCb experiment at $\sqrt{s} = 13~\mathrm{TeV}$, corresponding to an integrated luminosity of $5.4~\mathrm{fb^{-1}}$. The data cannot be fully explained by $B^{+} \to \psi(2S) K^{*+}$ contributions alone. A significantly better description of the data is obtained by adding a $T_{c\bar{c}}^{+}$ contribution decaying to $\psi(2S)\pi^{+}$. The properties of the $T_{c\bar{c}}^{+}$ structure are consistent with the exotic state $T_{c\bar{c}1}(4430)^{+}$ reported in the isospin-related $\bar{B}^{0} \to \psi(2S) K^{-} \pi^{+}$ decay. Effects of a possible $T_{c\bar{c}1}(4430)^{+} \to \bar{D}_{1}^{*}(2600)^{0} D^{+}$ decay mode on the $T_{c\bar{c}1}(4430)^{+} \to \psi(2S)\pi^{+}$ mass distribution are investigated through a Flatté parametrization, providing constraints on the relative decay strength. A description of the $T_{c\bar{c}1}(4430)^{+}$ structure using the triangle singularity mechanism is studied and also found to be consistent with the data.

\end{abstract}

\vspace*{1.cm}

\begin{center}
  Published in Physical Review D 113 (2026) L071101

\end{center}

\vspace{\fill}

{\footnotesize 
\centerline{\copyright~\papercopyright. \href{\paperlicenceurl}{\paperlicence}.}}
\vspace*{2mm}

\end{titlepage}


\newpage
\setcounter{page}{2}
\mbox{~}
%
%
%
%

%% file: body.tex
Hadrons beyond conventional $q\bar{q}$ mesons or $qqq$ baryons have been anticipated since the introduction of the quark model~\cite{GellMann:1964nj}. 
Referred to as exotic states, such hadrons serve as distinctive showcases of the intricate nonperturbative nature of QCD at low energies. 
A wealth of exotic candidates has been observed experimentally, whose properties do not fit in the known conventions of hadron spectroscopy~\cite{Guo:2017jvc,Chen:2022asf,Liu:2019zoy,Olsen:2017bmm,Johnson:2024omq,Husken:2024rdk,wang2025reviewqcdsumrules}.
The first evidence for a charged charmonium-like structure was found in the $\psitwos \pip$ final state of the $\decay{B}{\psitwos K \pip}$ decay~\cite{Belle:2007hrb,BelleZ4430}.\footnote{Unless otherwise specified, charge-conjugated states or decays are implied throughout this Letter.} 
The minimum quark content of the observed structure, referred to as $\myZcp$ in the following, is $c\bar{c}u\bar{d}$, and its spin-parity has been determined by an amplitude analysis to be $J^P = 1^+$ 
~\cite{Belle:2013shl,LHCb-PAPER-2014-014,Beiter:2023ltc}.


Many theoretical interpretations of the $\myZcp$ structure using kinematic effects or dynamical models have been  proposed. 
In dynamical interpretations, the $\myZcp$ structure is regarded as a genuine exotic state,  
described as a hadronic molecule with meson-meson interactions\cite{Meng:2007fu,Liu:2007bf,Ding:2008mp} or a compact tetraquark formed by color-exchange interactions~\cite{Maiani:2014aja,Deng:2015lca,Wang:2014vha,PhysRevD.96.034026}. 
However, these dynamical interpretations face challenges in explaining the $\myZcp$ structure. 
In the hadronic molecular scenario, the $\myZcp$ structure is interpreted as an S-wave $\Dbar^*D_1$ hadronic molecule with possible spin-parity quantum numbers $J^P = 0^-, 1^-$ or $2^-$~\cite{Meng:2007fu,Liu:2007bf,Ding:2008mp}, which is strongly disfavored by the LHCb analysis reported in Ref.~\cite{LHCb-PAPER-2014-014}. 
In the compact tetraquark scenario, an octet group of meson states is predicted; however none of these, apart from the possible $\myZcp$ state, have been observed. 
In contrast, in the kinematical interpretations, the $\myZcp$ structure is generated as a kinematical singularity in the $B$-decay amplitude~\cite{Coleman:1965xm,Guo:2019twa}. 
In one possible model~\cite{Nakamura:2019btl}, the rescattering of $\psi(4230)\pip$ hadrons in the $\Bzb\to \Kstarb(892)^0\psi(4230)$, $\Kstarb(892)^0\to\Km\pip$ cascade decay could form a peaking structure in the $\psitwos\pip$ final state, consistent with the measured $\myZcp$ properties. 
Additionally, lattice QCD simulations cannot predict the $\myZcp$ state due to the entanglement of many possible decay channels~\cite{Bicudo:2022cqi}.
To date, there is no consensus on the nature of $\myZcp$ structure.
Further experimental and theoretical studies are needed for new insights.

This Letter reports on an amplitude analysis performed on the $\myDecay$ decay, analogous to the isospin-related decay channel $\decay{\Bzb}{\psitwos \Km \pip}$.
The data used are proton-proton ($pp$) collisions recorded by the LHCb experiment, at a center-of-mass energy of $13\tev$, and corresponding to an integrated luminosity of 5.4\invfb. 
The \lhcb detector is a single-arm forward spectrometer covering the pseudorapidity range $2 < \eta < 5$, described in detail in Refs.~\cite{LHCb-DP-2008-001,LHCb-DP-2014-002}.
Candidate $\myDecay$ decays are formed by combining $\pip$ candidate tracks with $\psitwos$ and $\KS$ meson candidates reconstructed in the $\psitwos \to \mup \mun $ and $\KS \to \pip \pim$ decay modes, respectively. 
Particle identification, track quality, and impact parameter requirements are applied to all final-state particles to ensure consistency with the signal decay.
The reconstructed $\Bp$, $\psitwos$  and $\KS$ mesons are required to have good vertex-fit $\chi^2$ and invariant masses close to their known values~\cite{PDG2024}. The background, dominated by random combinations of $\psitwos$, $\pip$, and $\KS$ candidates,
is further suppressed by a boosted decision tree~(BDT)~\cite{Breiman,AdaBoost} classifier. The classifier is implemented with the TMVA toolkit~\cite{Hocker:2007ht,TMVA4} and uses as discriminating variables the transverse momenta, vertex-fit quality, and particle-identification information of the final-state particles.

A simulated sample of $\myDecay$ decays, generated with the software packages
described in Refs.~\cite{Sjostrand:2007gs, Lange:2001uf, Allison:2006ve}, is used to model the effects of the detector acceptance and the imposed selection requirements.
The simulated decays are subjected to the same reconstruction and selection procedures as the data. 

The $\Bp$ candidate invariant-mass distribution, calculated with the $\psitwos$ and $\KS$ masses constrained to their known values~\cite{PDG2024},  is shown in Fig.~\ref{fig:fitmass} of the supplemental material~\cite{supplemental}. 
The signal yield is found to be $9600\pm100$, determined from an unbinned extended maximum-likelihood fit to the $\Bp$ mass spectrum, where the signal component is modeled by a modified Gaussian function~\cite{Skwarnicki:1986xj} with power-law tails on both sides, and the combinatorial background is described by an exponential function. 
The fit result is further used to assign a signal weight~\cite{Pivk:2004ty,DEMBINSKI2022167270} to each candidate, which is employed to perform background subtraction in the subsequent analysis steps following the sFit technique~\cite{Xie:2009rka,Langenbruch:2019nwe,DEMBINSKI2022167270}.

An amplitude analysis is performed to investigate the various contributions in the $\myDecay$ decay, where the amplitude models are developed following the helicity formalism~\cite{JACOB1959404}. 
The first model includes only $\Bp\to\psitwos\Kstarp$ contributions, with  excited $\Kstarp$ mesons decaying into $\KS \pip$. 
Four independent variables are needed to describe the kinematics of the cascade decay, referred to as the $\Kstar$ chain: $\decay{\Bp}{\psitwos\Kstarp }$, $\decay{\Kstarp}{\KS \pip}$, $\decay{\psitwos}{\mu^+ \mu^-}$. The variables are chosen in the fit to be the $\Kstarp$ invariant mass, $m_{K\pi}$, the cosines of the helicity angles of the $\Kstar$ and $\psitwos$ decays, $\cos \theta_{\Kstar}$ and $\cos \theta_{\psi}$, and the angle between the $\Kstarp$ and $\psitwos$ decay planes $\phi$, as defined in Fig.~\ref{Kstchain} of the supplemental material~\cite{supplemental}. The four independent variables are calculated with the $\Bp$ and $\psitwos$ masses constrained to their known values~\cite{PDG2024}.

The total amplitude model is constructed using the isobar approach~\cite{PhysRev.105.1874}, where a coherent sum is taken over various $\Kstarp$ resonances and an incoherent sum over final-state $\mup$ and $\mun$ helicities. 
All known $\Kstarp$ resonances with masses below the upper kinematic limit ($1.593\gevcc$) are considered in the baseline amplitude model. These comprise the $K^*_0(700)^+$ and $K^*_0(1430)^+$ mesons with $J=0$, the $K^*(892)^+$ and $K^*(1410)^+$ mesons with $J=1$, and the $K^*_2(1430)^+$ meson with $J=2$.
The masses and widths of these $\Kstarp$ resonances are fixed to their known values~\cite{PDG2024}, except for the dominant $K^*(892)^{+}$ state, whose mass and width are allowed to float in the amplitude fit with Gaussian constraints to the known values~\cite{PDG2024}. 
The amplitude is constructed using $LS$ (orbital angular momentum and spin coupling) bases instead of helicity bases~\cite{LHCB-PAPER-2015-029}. The relative momentum between the $\psitwos$ and $\Kstarp$ meson in the $\Bp$ rest frame is rather small, so only the amplitudes with the lowest possible orbital angular momentum  between $\psitwos$ and $\Kstarp$ mesons are considered in the baseline fit,  except for $\Bp\to\psitwos K^*(892)^{+}$ amplitudes where all three possible contributions are included. The effects of including amplitudes with higher orbital angular momenta are considered as systematic uncertainties.
Each amplitude is associated with a complex coupling, which is determined by the fit to data. 

The $\Kstarp$ resonances are modeled using relativistic Breit--Wigner amplitudes,
\begin{equation}
    BW(m_{K\pi}\lvert m_{0},\Gamma_{0}) = \frac{1}{m_0^2-m_{K\pi}^2-im_0\Gamma(m_{K\pi})},
\end{equation}
where  \mbox{$\Gamma(m_{K\pi}) = \Gamma_0\left(q/q_0\right)^{2L+1} \left(m_0/m_{K\pi}\right)B^2_{L}(q,q_0,d)$} is the mass-dependent width, and $m_0$ and $\Gamma_0$ are the mass and natural width of a $\Kstarp$ resonance.
The variable $q$ refers to the momentum of the $\KS$ meson in the $\Kstarp$ rest frame, and $q_0$ denotes the value evaluated at the resonance peak $m_{K\pi} = m_0$. 
The orbital angular momentum between $\KS$ and $\pip$ mesons, $L$, is fixed by the $\Kstarp$ resonance spin.   
The Blatt–Weisskopf form factor is also applied to the relativistic Breit–Wigner amplitude, together with an orbital angular momentum suppression factor, to account for the effects of higher partial waves with nonzero orbital angular momentum. In the  Blatt--Weisskopf form factor~\cite{Blatt:1952ije}, $B_{L}(q,q_0,d)$, the radius $d = 3~(\!\gevc)^{-1}$ is used for intermediate states  and $d = 5~(\!\gevc)^{-1}$ for the $\Bp$ meson~\cite{LHCb-PAPER-2014-014}.

A weighted, unbinned maximum-likelihood fit is performed to the data in four kinematic variables, following the sFit technique~\cite{Xie:2009rka,Langenbruch:2019nwe,DEMBINSKI2022167270}, 
which removes the need of an explicit description of the background in the fit. 
In the fit, the efficiency dependence on kinematic variables is determined using simulated samples of the \myDecay decay.

\begin{figure}[!t]
\begin{center}
\includegraphics[width=0.32\columnwidth]{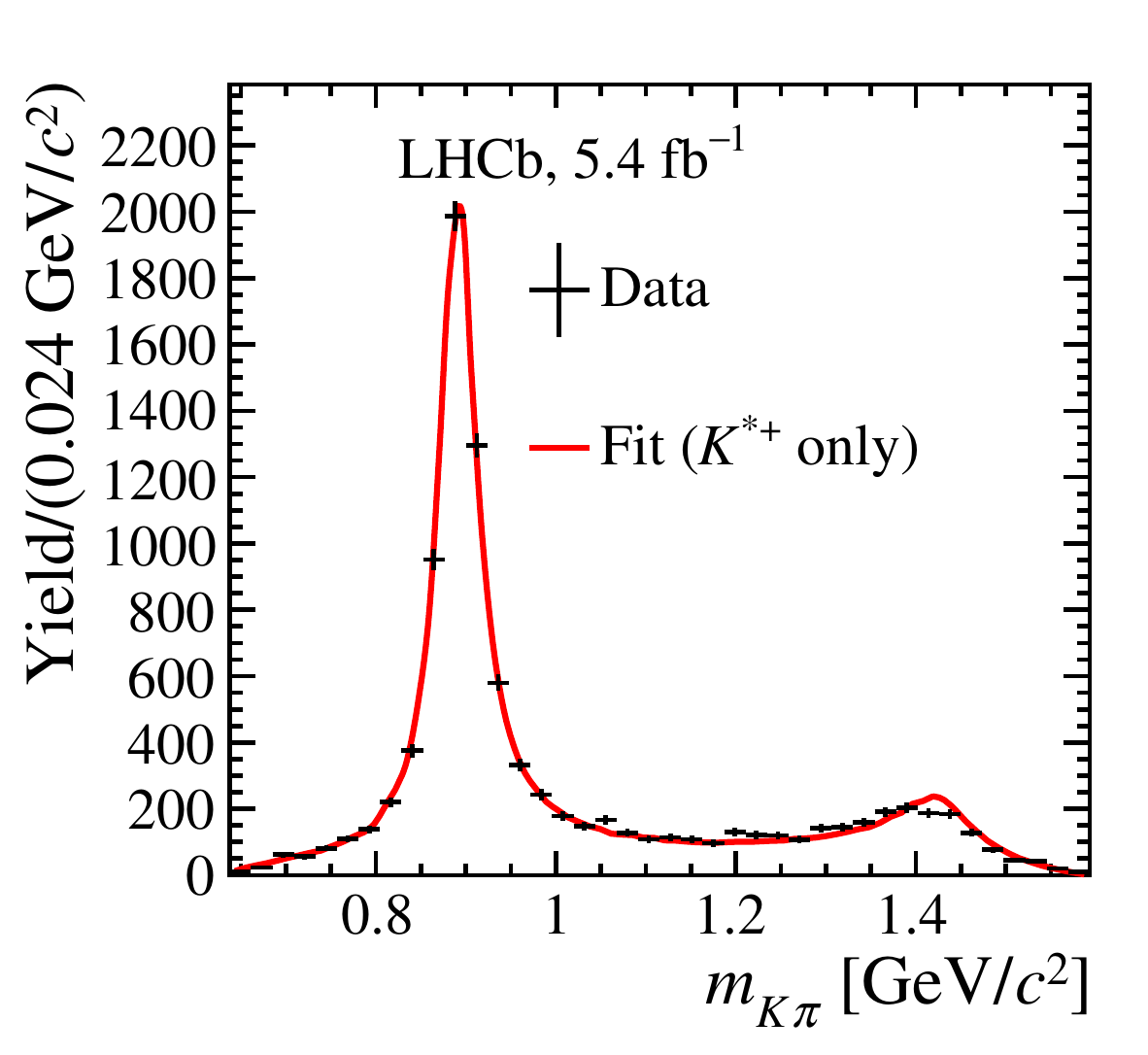}
\includegraphics[width=0.32\columnwidth]{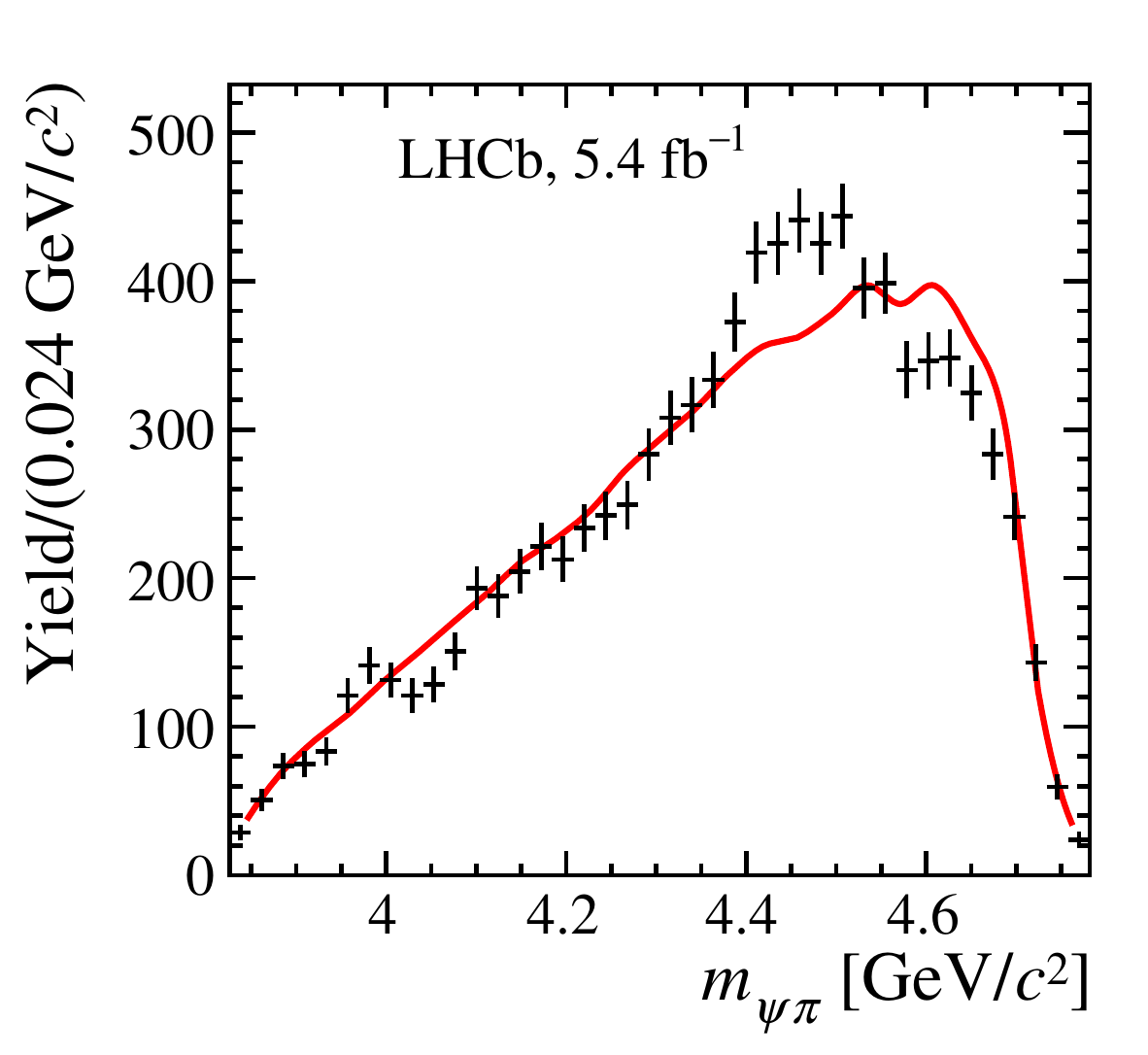}
\includegraphics[width=0.32\columnwidth]{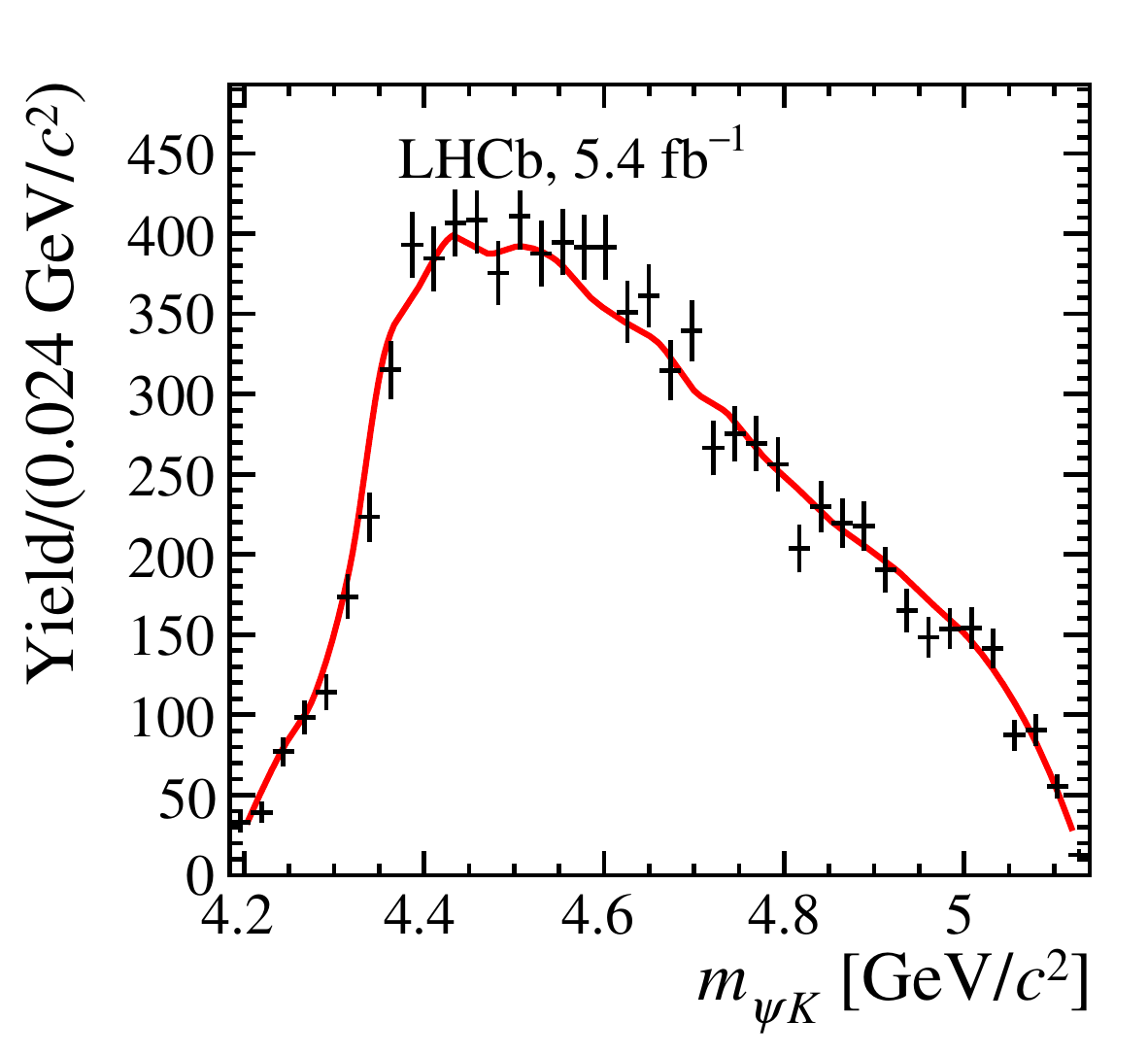}
\end{center}
\caption{Invariant-mass distributions of (left) the $K^{0}_{\mathrm{S}}\pi^{+}$, (middle) $\psi(2S)\pi^{+}$ and (right) $\psi(2S)K^{0}_{\mathrm{S}}$ pairs for background-subtracted data (black dots), together with projections of the results of an amplitude fit (red solid line) with only $B^{+} \to \psi(2S)K^{*+}$ contributions.
}
\label{fig:Null}
\end{figure}

The background-subtracted invariant-mass distributions for $\KS\pip$ ($m_{K\pi}$),  $\psitwos\pip$ ($m_{\psi \pi}$)  and $\psitwos\KS$ ($m_{\psi K}$) pairs are shown in Fig.~\ref{fig:Null}, together with the results of the fit considering only $\mbox{\Bp\to \psitwos\Kstarp}$ contributions.
While the $\Kstarp$ model can describe the $m_{K\pi}$ and $m_{\psi K}$ distributions well, the fit does not provide an adequate description of the $m_{\psi \pi}$ shape, especially in the range  $4.2<m_{\psi \pi}<4.7\gevcc$.
Adding more $\Kstarp$ resonances or nonresonant $\KS\pip$ contributions does not significantly improve the fit quality in this region.

\begin{figure}[htb!]
\begin{center}
\includegraphics[width=0.48\columnwidth]{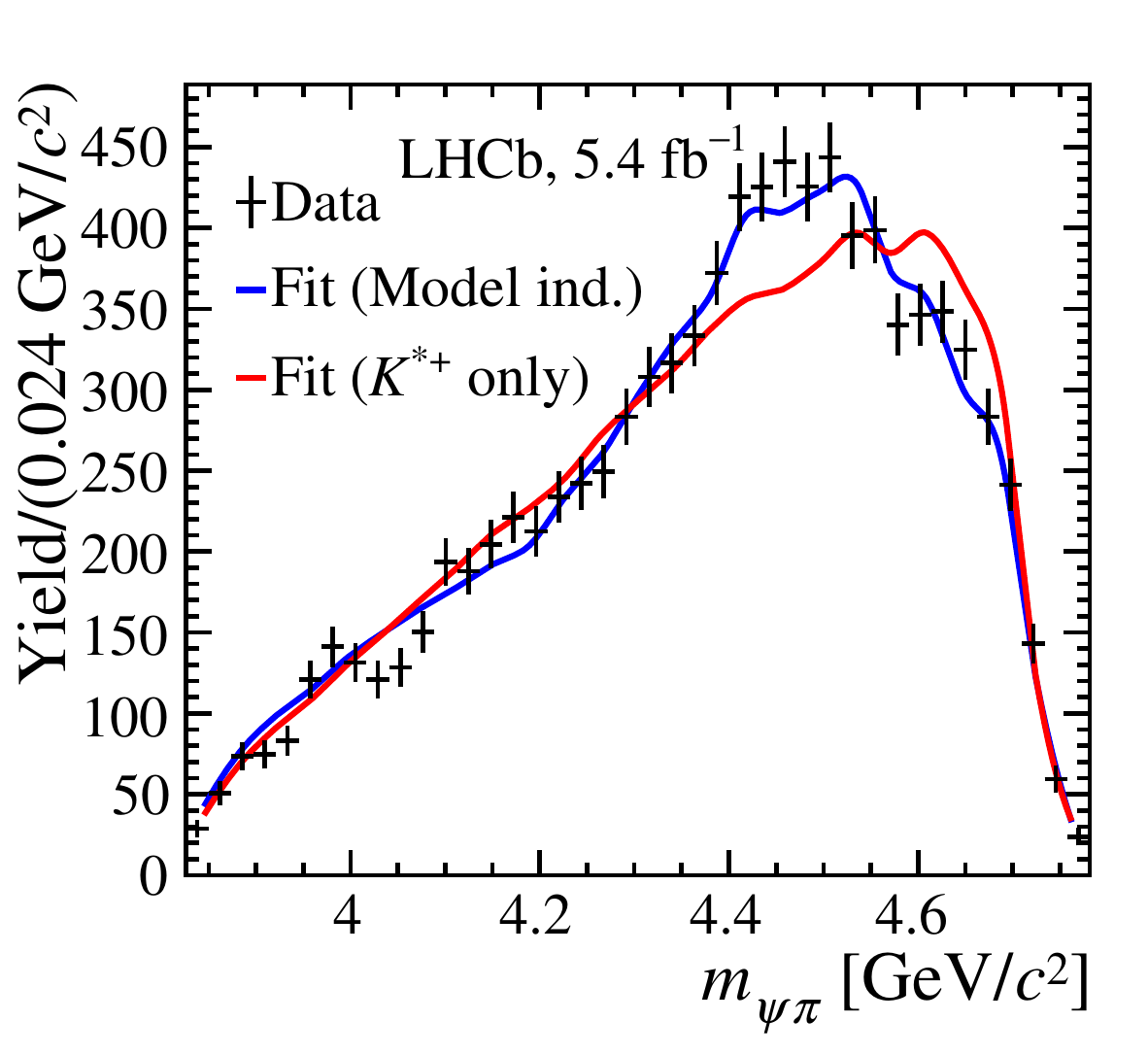}
\includegraphics[width=0.48\columnwidth]{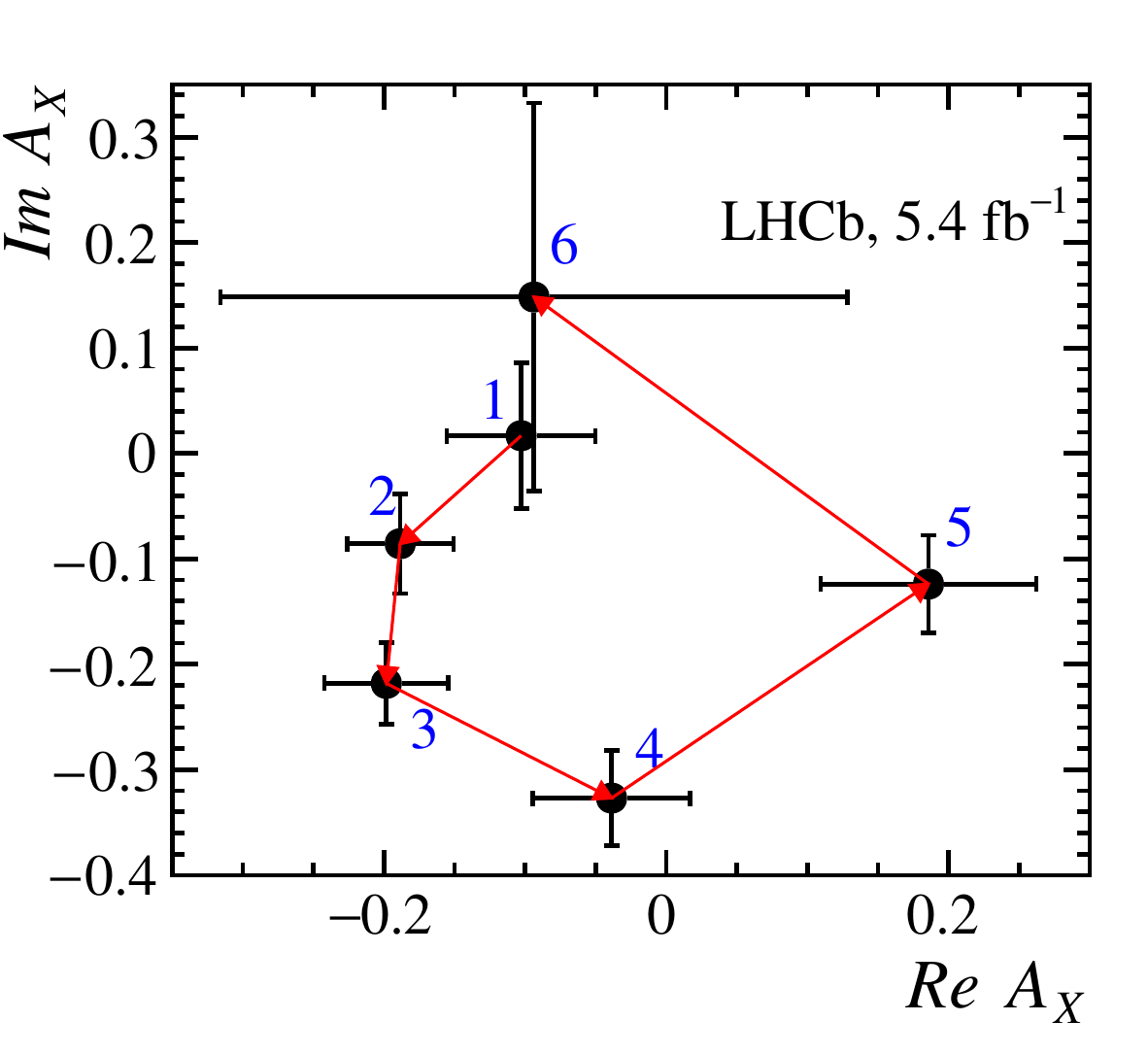}
\end{center}
\caption{(Left) Distribution of the $\psi(2S)\pi^+$ invariant-mass  of background-subtracted data. The projections of the fit including (red) only $K^{*+}$ resonances and (blue) a model-independent amplitude are also shown. 
(Right) Argand diagram for the amplitude $A_{X}$, showing the complex amplitude values at six points. Each point corresponds to a different value of $m_{\psi\pi}$, which increases in the counterclockwise direction.}
\label{fig:model-independent}
\end{figure}

The unidentified structure in the $m_{\psi \pi}$ spectrum (in the following referred to as structure $X$) is investigated by adding a $\psitwos \pip$ component to the $\Kstarp$ amplitudes. In the coherent sum of the $\Kstarp$ and $\psitwos \pip$ amplitudes, the muon helicity reference frames are aligned by applying a rotation that depends on the four independent variables $m_{K\pi}$, $\cos \theta_{\Kstar}$, $\cos \theta_{\psi}$ and $\phi$~\cite{LHCB-PAPER-2015-029}.
A model-independent approach is employed to describe the $X$ invariant-mass distribution, with a cubic spline interpolation of the amplitudes at six fixed  $m_{\psi\pi}$ values chosen equidistantly in the range $[4.2, 4.7]\gevcc$.
The complex amplitudes at six fixed  $m_{\psi\pi}$ values, with
the orbital angular momentum between $\psitwos$ and $\pip$ fixed to zero, describe the data well.
The projection of the fit on the $m_{\psi\pi}$ distribution is shown on the left side of Fig.~\ref{fig:model-independent}.
The corresponding amplitudes in the complex plane (the Argand diagram~\cite{osti_4492200}) are presented on the right side of Fig.~\ref{fig:model-independent}, revealing a circular phase shift as a function of $m_{\psi\pi}$ from lower to higher $\psitwos\pip$ masses, suggesting that the $X$ structure is unlikely to be caused by statistical fluctuations in data. Counter-clockwise evolution with mass is consistent with a resonant behavior  but could also originate from other physical effects, 
such as triangle singularity~\cite{Guo:2019twa, Nakamura:2019btl}.

\begin{figure}[tb]
\begin{center}
\includegraphics[width=0.45\columnwidth]{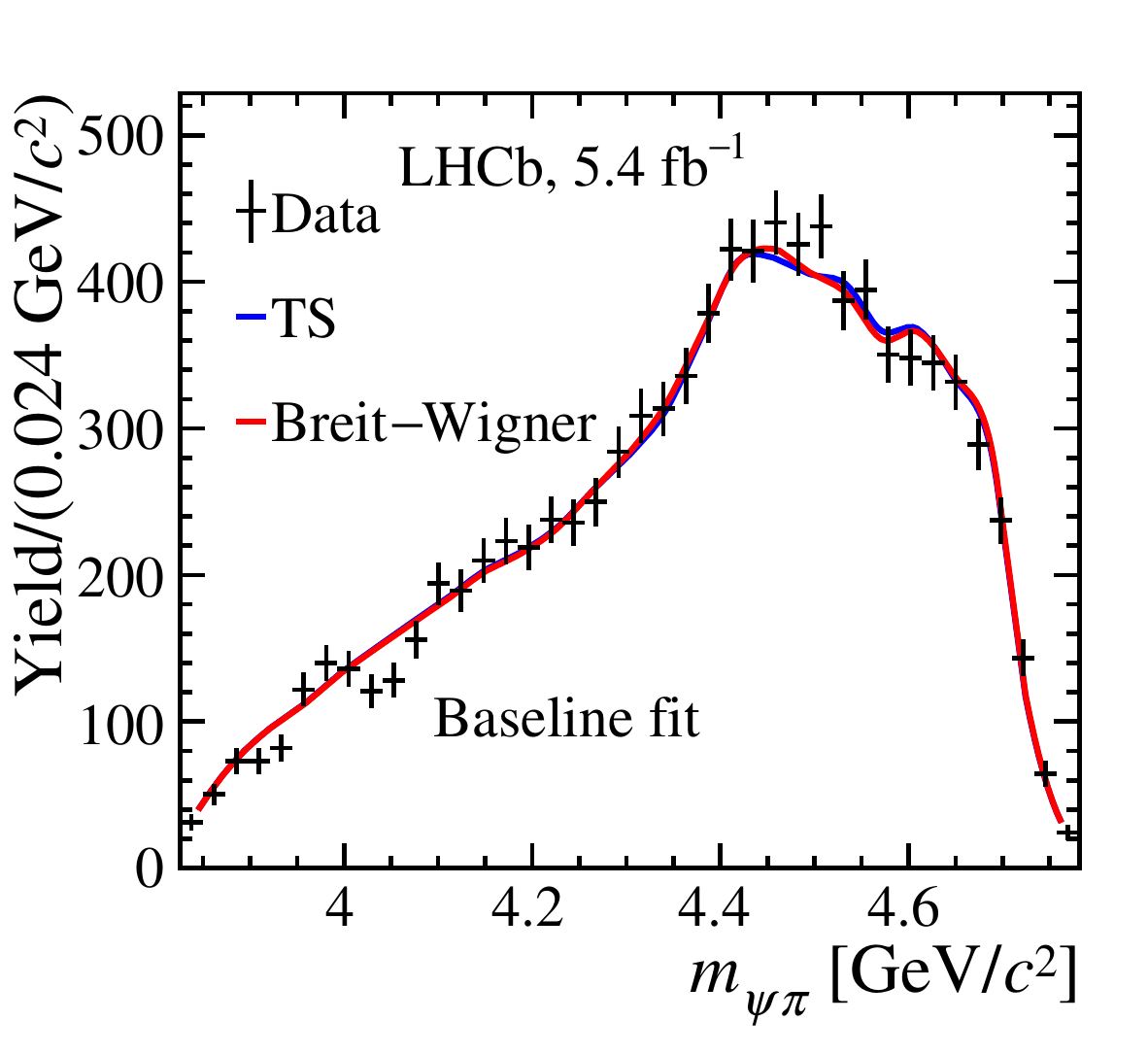}
\includegraphics[width=0.45\columnwidth]{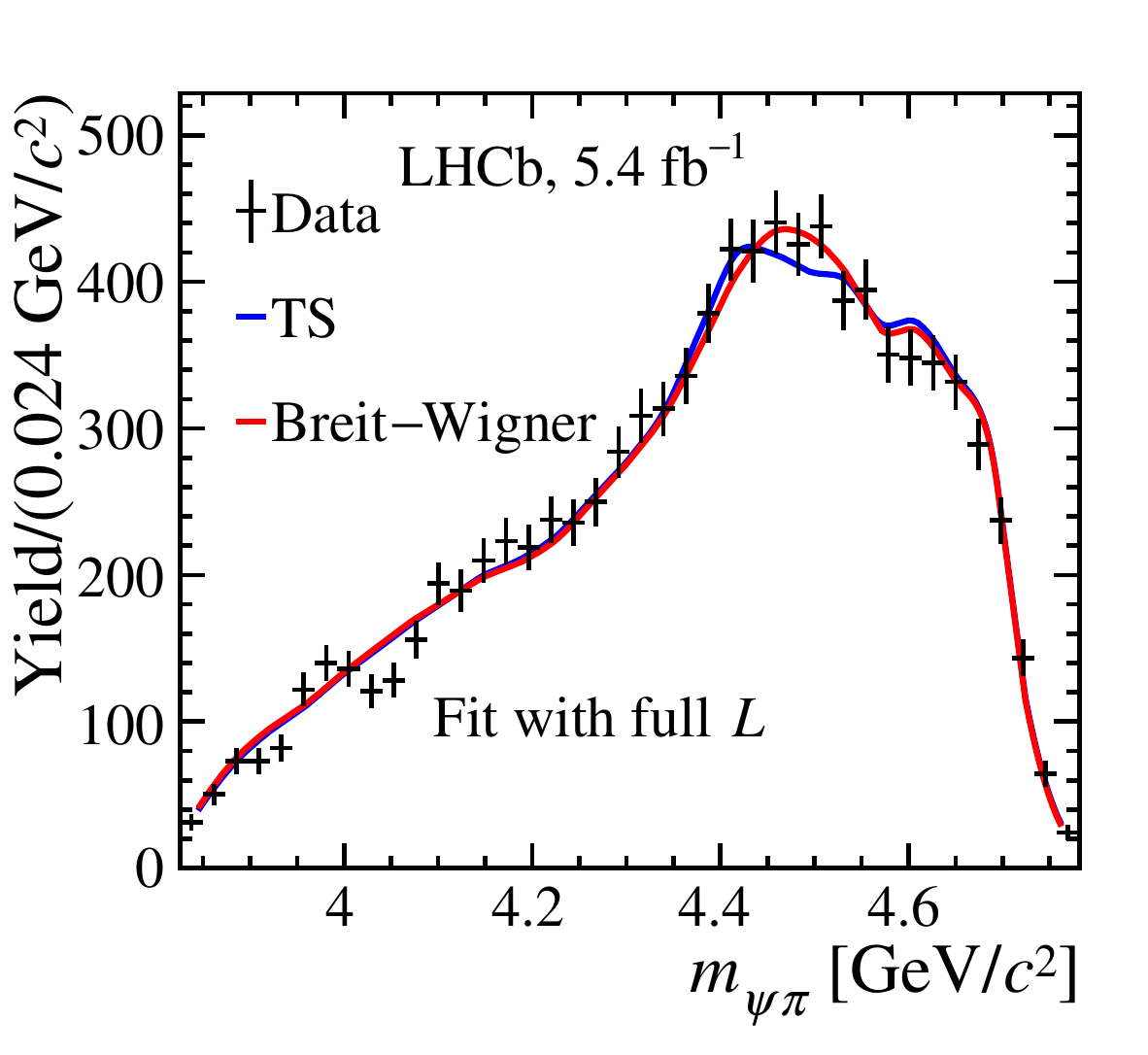}

\end{center}
\caption{Distribution of the $\psi(2S)\pi$ invariant-mass of background-subtracted data, with the projections of the fit including the relativistic Breit--Wigner parametrization (red) or the triangle singularity amplitude (blue) for the (left) baseline fit and (right) the fit with full orbital angular momenta.}
\label{fig:TS}
\end{figure}

Model-dependent approaches are used to extract the properties of the $X$ structure.
In the first model the structure is considered as a resonance, $\myZc$, decaying into the $\psitwos\pip$ final state and with the mass distribution described by a relativistic Breit--Wigner function.
The mass and width of the $\myZc$ state are allowed to float, and only the lowest possible orbital angular momentum between the $\psitwos$ and $\pip$ mesons is considered in the baseline fit.
A satisfactory description of data is achieved for the spin-parity $J^P (\myZc) = 1^+$. 
The projection of the amplitude fit result onto the $\psitwos \pip$ invariant mass is shown by the red curve in Fig.~\ref{fig:TS}.
The $\myZc$ mass and width are measured to be  $M_{\myZc} = 4.452\pm0.016^{+0.055}_{-0.033} \gevcc$ and $\Gamma_{\myZc} = 0.174\pm0.019^{+0.083}_{-0.020} \gev$. The $\myZc$ fit fraction in the $\myDecay$ decay is determined to be $f_{\myZc} = (3.7\pm0.6^{+4.0}_{-0.7})\%$, where the first uncertainty is statistical and the second systematic. 
The fit fraction of a specific contribution is calculated as the integral of its amplitude squared over the full phase space divided by that of the total amplitude squared. 
The sources of systematic uncertainties are described below. 
The  $\myZc$ properties are consistent with those of the $\myZcp$ structure observed in the $\decay{\Bzb}{\psitwos \Km\pip}$ decay~\cite{PDG2024}. 
The significance of the $\myZc$ state is evaluated based on $2\Delta\ln L = 2\ln L_{\myZc} - 2 \ln L_{\Kstarp}$ , where $L_{\Kstarp}$ and $L_{\myZc}$ refer to the likelihood functions without and with the $\myZc$ component, each evaluated at its maximum.
The quantity $2\Delta\ln L$ follows a $\chi^2$ distribution, with a number of degrees of freedom approximately twice  
the number of additional free parameters in the $\myZc$ fit compared to the  $\Kstarp$-only fit, after accounting for the look-elsewhere effect~\cite{LHCb-PAPER-2014-014,LHCb-PAPER-2015-038}.
The statistical significance of the $\myZc$ state with $J^P=1^+$ is determined to be more than $16\sigma$, and remains above $9\sigma$ after including systematic effects, which is obtained by the fit with the smallest $\Delta\ln L$ among all systematic sources.

The $J^P$ assignment for the $\myZc$ state is determined by  comparing the likelihood values of the fits with different hypotheses. The $0^+$ assignment is excluded due to parity conservation.
The hypotheses $0^-$, $1^-$, $2^-$, $2^+$ are rejected relative to the $1^+$ hypothesis with a significance of more than $6\sigma$, $11\sigma$, $7\sigma$, $11\sigma$, respectively, after accounting for the systematic uncertainty. 
Therefore, the $\myZc$ spin-parity is unambiguously determined as $J^P=1^+$, consistent with the quantum numbers of the $\myZcp$ state observed in the $\decay{\Bzb}{\psitwos \Km\pip}$ decay~\cite{PDG2024}. 
Consequently, it is reasonable to conclude that the $\myZc$ state in the $\myDecay$ decay corresponds to that observed in the $\decay{\Bzb}{\psitwos \Km\pip}$ decay~\cite{Belle:2013shl,LHCb-PAPER-2014-014,LHCb-PAPER-2015-038}, being produced by the decay processes related via isospin symmetry.

Various sources of systematic uncertainties are studied for the mass, width and fit fraction measurements of the $\myZc$ state, including 
varying the masses and widths of $\Kstarp$ resonances; 
adding a nonresonant $\KS \pip$ component with zero spin; 
including the low mass tails of $K^*(1680)^+$ and $K^*_3(1780)^+$ into the amplitude fit; 
parameterizing the $K^*_0(700)^+$ and $K^*_0(1430)^+$ states with the LASS model~\cite{ASTON1988493};
varying the Blatt–Weisskopf radius for both $B^+$ and intermediate states' decays between $3$ and $5~(\!\gevc)^{-1}$; 
allowing higher orbital angular momenta to contribute; 
extracting the signal projection weights using alternative signal or background models in the $\Bp$ invariant-mass fit.
The differences between the results of the baseline fit and alternative fits are taken as the systematic uncertainties. 
Based on previous measurements of $\psi(nS)\pi^+$ spectra~\cite{LHCb-PAPER-2014-014,LHCb-PAPER-2015-038,Belle:2013shl, Beiter:2023ltc}, an additional $T_{c\bar{c}1}(4200)^+\to\psitwos\pip$ or $T_{c\bar{c}0}(4240)^+\to\psitwos\pip$ component is added to the amplitude with the $T_{c\bar{c}1}(4200)^+$ or $T_{c\bar{c}0}(4240)^+$ parameters fixed to their known values~\cite{PDG2024} as another source of systematic uncertainty. The significance of either the $T_{c\bar{c}}(4200)^+$ state or the $T_{c\bar{c}}(4240)^+$ state is determined to be $4\sigma$. 
Among systematic uncertainties related to the $\Kstarp$ modeling, which are not independent, only the maximum difference is retained and is combined with other sources to obtain the total systematic uncertainty.

In the model described above, the $\myZcp$ state is assumed to decay only into the $\psitwos\pip$ final state. However, if the $\myZcp$ has a molecular nature, it is expected to couple strongly to open-charm hadrons whose invariant-mass threshold lies near its mass.
Notably, the $\myZcp$ mass is close to the $\mydd$ production threshold, and its spin-parity is consistent with an S-wave $\mydd$ configuration.
To account for this possibility, the amplitude model is modified, taking into account the effect that the opening of the $\myZcp \to \mydd$ decay channel would have on the $\myZcp \to \psitwos\pip$ lineshape.
Within this framework, the $\myZcp$ resonance is modeled using the Flatté parametrization~\cite{Flatte:1976xu},
\begin{equation}
\label{FL}
F = \frac{1}{m_f^2 - m^2 - i (\rho_1 g_1^2 + \rho_2 g_2^2)},
\end{equation}
where the positive parameters $g_1$ and $g_2$ represent the coupling strengths to the $\psitwos\pip$ and $\mydd$ channels, respectively, and $\rho_1$ and $\rho_2$ are the corresponding phase-space factors, discussed further in the supplemental material~\cite{supplemental}.
The Flatt\'{e} parametrization reduces to the relativistic Breit--Wigner parametrization for $g_2=0 $. 
The amplitude fit with the $\myZcp$ state modeled with the Flatt\'{e} parametrization yields 
\mbox{$m_f = 4.452\pm0.022^{+0.103}_{-0.005} \gevcc$}, 
\mbox{$g_1 = 1.58\pm0.17_{-0.82}^{+0.05} \gevcc$}, 
\mbox{$g_2 = 0.00\pm1.78\pm2.81 \gevcc$} 
and the fit fraction \mbox{$f = (3.7\pm0.6^{+3.7}_{-0.7}) \%$}.
The upper limit for the relative decay strength $R\equiv |g_2/g_1|$ is determined to be $R<6.8$ at the $95\%$ confidence level, using a profile-likelihood scan. 
This constrains the coupling of $\myZcp$ to the $\mydd$ final state.

Apart from dynamical interpretations, the origin of the $X$ structure could be kinematical. 
Following the model in Ref.~\cite{Nakamura:2019btl}, a triangle singularity mechanism is tested, where the $X$ structure in the $\psitwos\pip$ mass spectrum arises from the $\psi(4230)\pip\to \psitwos\pip$ rescattering in the $\decay{\Bp}{\psi(4230)\Kstar(892)^+}$, $\Kstar(892)^+\to\KS\pip$ cascade decay, as shown in Fig.~\ref{fig:tsdiagram} of the supplemental material~\cite{supplemental}. In such a triangle singularity mechanism, a relatively large $\decay{\Bp}{\psi(4230)\Kstar(892)^+}$ branching fraction is implied.
The $\Kstar(892)^+\psi(4230)\pip$ triangle diagram develops a singularity in the S-matrix of $\psi(4230)\pip\to\psitwos\pip$ transition when the intermediate states are simultaneously on shell.
The amplitude for the triangle singularity is obtained through integration over the triangle diagram, and the $\psitwos\pip$ invariant-mass distribution is determined by the properties of the involved intermediate and final-state hadrons, leaving no free parameters apart from an overall complex coupling.
The triangle singularity also exhibits a phase shift behavior as a function of the $\psitwos\pip$ invariant mass, very similar to that of the Breit--Wigner distribution.
The result of the fit using the $\Kstar(892)^+\psi(4230)\pip$ triangle amplitude to model the $X$ structure is shown in Fig.~\ref{fig:TS} projected onto the $\psitwos\pip$ invariant mass. The model provides a reasonable description of the data.
The fit yields a fit fraction $f_{X} = (3.9\pm0.7^{+3.3}_{-0.1})  \%$, consistent with that of the fit using the Breit--Wigner function.
 In some of the scenarios considered in the study of systematic uncertainties, the fit quality of the triangle contribution is reduced relative to that of the Breit--Wigner lineshape.
%
For example, including higher orbital angular momenta between $\psitwos$ and $\Kstarp$ mesons, the quality of the fit with the triangle amplitude is slightly worse than that with the Breit--Wigner lineshape for the $X$ structure, as shown on the right of Fig.~\ref{fig:TS}. Larger samples may help to distinguish the two models. An alternative triangle singularity model, described in
Ref.~\cite{Gribov:2009cfk}, was also investigated. It features longer tails in the $m_{\psi\pi}$ distribution, and cannot provide a satisfactory description of data.

To summarize, a full amplitude analysis is performed to the $\myDecay$ decay using $pp$ collision data collected by the~\lhcb experiment at a center-of-mass energy of $ 13\tev$ and corresponding to an integrated luminosity of $5.4\invfb$.
With contributions of known $\Kstarp$ resonances only, a discrepancy between data and the amplitude fit is observed, most obvious in the $\psitwos\pip$ invariant-mass distribution around $4.5\gevcc$. 
The discrepancy is resolved by including a component in the  $\psitwos\pip$ final state in the amplitude. 
A model-independent description of this component reveals a peaking structure with its complex phase evolving as a function of the $\psitwos\pip$ invariant mass.
Modeling the structure with a Breit--Wigner function gives a measurement of its mass, width, spin-parity and fit fraction in $\myDecay$ decays of:
 $M_{\myZc}=4.452\pm0.016^{+0.055}_{-0.033}\gevcc$, $\Gamma_{\myZc}=0.174\pm0.019^{+0.083}_{-0.020}\gev$, $J^P=1^+$ and $f_{\myZc} = (3.7\pm0.6^{+4.0}_{-0.7}) \%$.
The results are consistent with the exotic candidate $\myZcp$ reported by the Belle and LHCb collaborations in the $\decay{\Bzb}{\psitwos \Km \pip}$ decay. 
An additional fit is performed with a formalism that includes the possible $\myZcp$ decay into the $\mydd$ final state. An upper limit is set on the coupling strength of the $\myZcp\to\mydd$ decay relative to that of the $\myZcp\to\psitwos\pip$ decay.
A reasonable description of the $\myZcp$ structure is also achieved using a kinematical model incorporating the singularity in the  $\psi(4230)\Kstarp\pip$ triangle diagram~\cite{Nakamura:2019btl}.\\
This analysis reports the observation of the $\myZcp$ structure in the $\myDecay$ decay, and is the first experimental investigation into the nature of the $\myZcp$ structure using a hadronic molecule-motivated model and the amplitude of a triangle diagram with a full amplitude analysis.
These results provide valuable insights into the nature of the $\myZcp$ structure.

%% file: acknowledgements.tex
\section*{Acknowledgements}
%
%
\noindent We express our gratitude to our colleagues in the CERN
accelerator departments for the excellent performance of the LHC. We
thank the technical and administrative staff at the LHCb
institutes.
We acknowledge support from CERN and from the national agencies:
ARC (Australia);
CAPES, CNPq, FAPERJ and FINEP (Brazil); 
MOST and NSFC (China); 
CNRS/IN2P3 (France); 
BMFTR, DFG and MPG (Germany);
INFN (Italy); 
NWO (Netherlands); 
MNiSW and NCN (Poland); 
MCID/IFA (Romania); 
MICIU and AEI (Spain);
SNSF and SER (Switzerland); 
NASU (Ukraine); 
STFC (United Kingdom); 
DOE NP and NSF (USA).
We acknowledge the computing resources that are provided by ARDC (Australia), 
CBPF (Brazil),
CERN, 
IHEP and LZU (China),
IN2P3 (France), 
KIT and DESY (Germany), 
INFN (Italy), 
SURF (Netherlands),
Polish WLCG (Poland),
IFIN-HH (Romania), 
PIC (Spain), CSCS (Switzerland), 
and GridPP (United Kingdom).
We are indebted to the communities behind the multiple open-source
software packages on which we depend.
Individual groups or members have received support from
Key Research Program of Frontier Sciences of CAS, CAS PIFI, CAS CCEPP, 
Fundamental Research Funds for the Central Universities,  and Sci.\ \& Tech.\ Program of Guangzhou (China);
Minciencias (Colombia);
EPLANET, Marie Sk\l{}odowska-Curie Actions, ERC and NextGenerationEU (European Union);
A*MIDEX, ANR, IPhU and Labex P2IO, and R\'{e}gion Auvergne-Rh\^{o}ne-Alpes (France);
Alexander-von-Humboldt Foundation (Germany);
ICSC (Italy); 
Severo Ochoa and Mar\'ia de Maeztu Units of Excellence, GVA, XuntaGal, GENCAT, InTalent-Inditex and Prog.~Atracci\'on Talento CM (Spain);
SRC (Sweden);
the Leverhulme Trust, the Royal Society and UKRI (United Kingdom).

%% file: supplementary-app.tex
\clearpage

\begin{center}
    \LARGE\textbf{Observation and investigation of the \myZcp structure in \myDecay decays} \\[1em]
    \large \textit{Supplemental material}
\end{center}

\vspace{2em}

\label{sec:Supplementary-App}

\section{\boldmath{$B^+$} invariant-mass distribution}

\begin{figure}[htb!]
\begin{center}
\includegraphics[width=0.8\columnwidth]{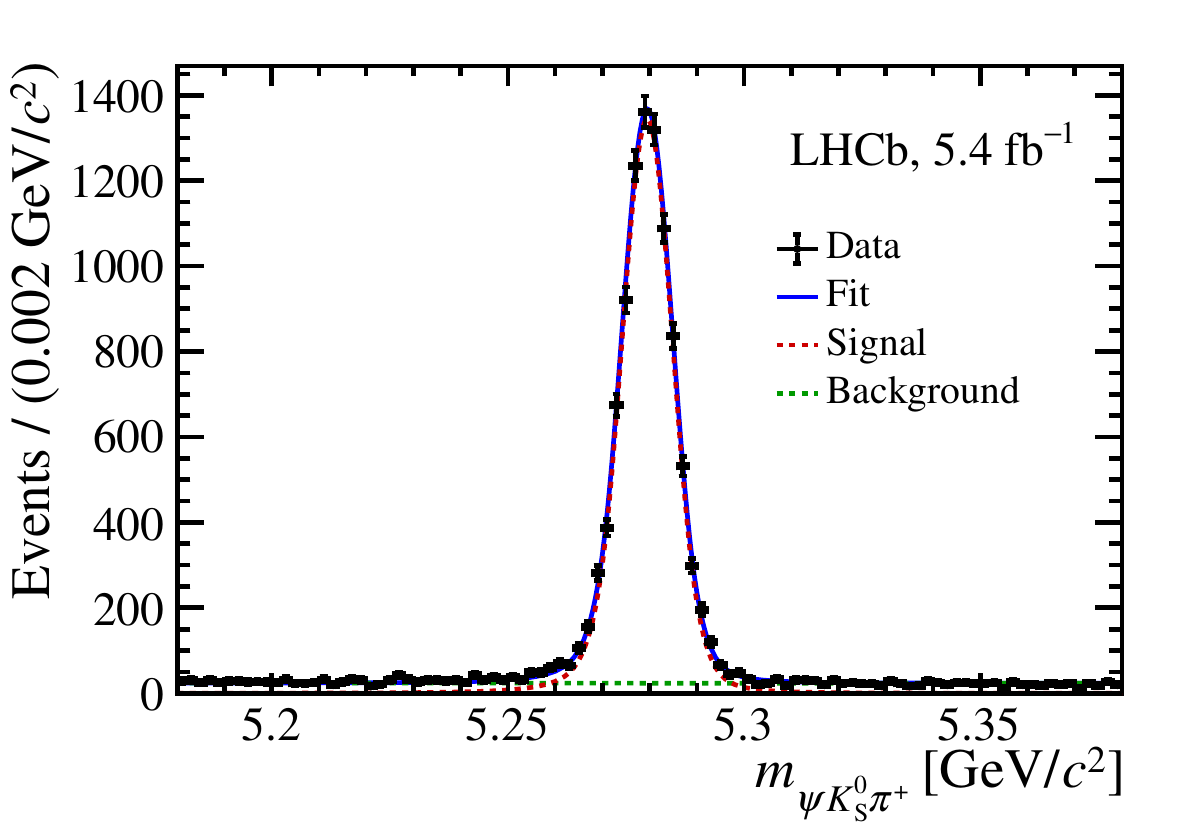}
\end{center}
\caption{Invariant-mass distribution of selected $B^{+} \to \psi(2S) K_{\text{S}}^{0} \pi^{+}$ candidate decays, with the fit result also shown. }
\label{fig:fitmass}
\end{figure}

The distribution of the $\Bp$ candidate invariant mass is shown in Fig.~\ref{fig:fitmass}.

\section{Kinematic variables used in the amplitude model}

\begin{figure}[!htbp]
    \centering
    \includegraphics[width = 0.96\columnwidth]{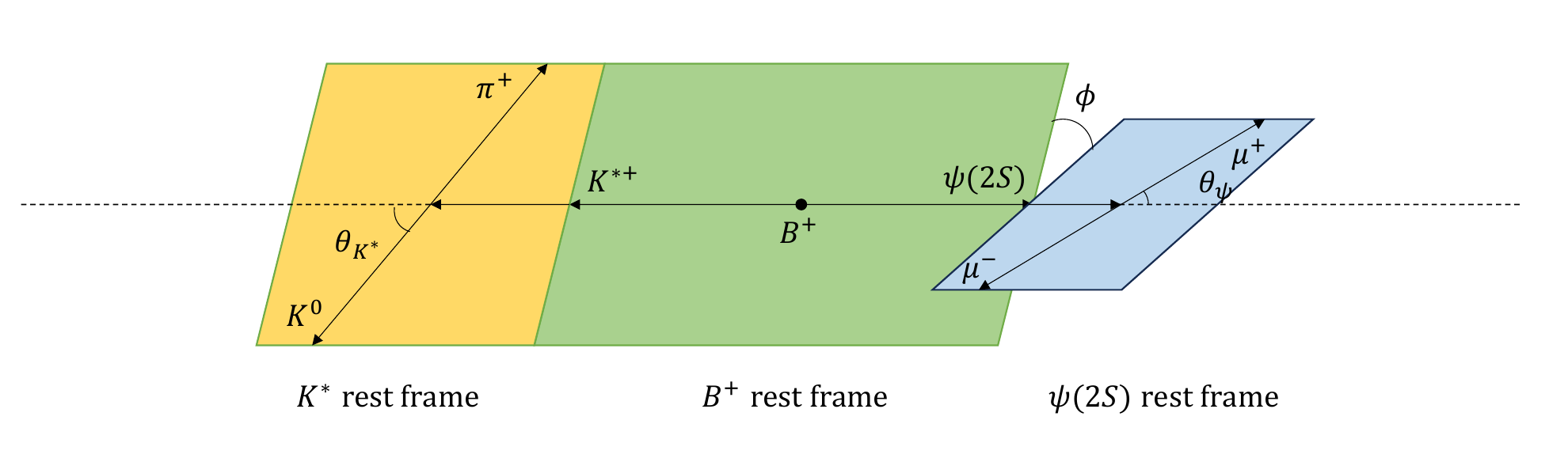}
\caption{The definition of the four independent angular observables used in the amplitude model.}
\label{Kstchain}
\end{figure}

The four independent variables used to describe the kinematics of the $\Kstarp$ decay chain are chosen to be
\begin{itemize}
    \item $m_{K\pi}$, invariant mass of the $\KS \pip$ system;
    \item $\cos{\theta_{\Kstar}}$, the helicity angle of the $\Kstarp$ decay, which is the angle between the momentum direction of the $\KS$ meson and the opposite momentum direction of the $\Bp $ meson in the $\Kstarp$ rest frame;
    \item $\cos{\theta_{\psi}}$, the helicity angle of the $\psi$ decay, which is the angle between the momentum direction of the $\mup$ lepton and the opposite momentum  direction of the $\Bp$ meson in the $\psitwos  $ rest frame;
    \item $\phi$, the angle between the $\Kstarp \rightarrow \KS\pip$ decay plane and the $\psitwos  \rightarrow\mup \mun$ decay plane.
\end{itemize}
The definitions of angular variables are sketched in Fig.~\ref{Kstchain}.

\section{Flatt\'{e} parametrization}

In the Flatt\'{e} amplitude described by Eq.~\ref{FL}, $\rho_1$ and $\rho_2$ denote the phase-space factors for the $\myZcp$ state decay into the $\psitwos\pip$ and $\mydd$ final states respectively. The phase-space factor is defined as $\rho_1 = 2q/m$, 
where $m$ is the $\psitwos \pip$ invariant mass and $q$ is the momentum of the $\pip$ meson in the \myZcp rest frame. 
The definition of the phase-space factor $\rho_2$ is more involved due to the non-negligible width of the \mydst meson. In this case, $\rho_2$ is taken as the three-body phase-space of the $\decay{\myZcp}{\Dp(\mydst \to \Dm \pip)}$ decay, where the \mydst is described by a relativistic Breit--Wigner function with parameters fixed to known values~\cite{PDG2024}. The real part of the corresponding three-body phase-space factor is given as
\begin{equation}
    \Real (\rho_2) = \frac{1}{\cal N} \int_{m_{D^{\scalebox{0.4}{\textnormal{--}}}} + m_{\pi^{\scalebox{0.4}{\textnormal{+}}}}}^{\,m - m_{D^{\scalebox{0.4}{\textnormal{+}}}}}\frac{p\, q}{m} |BW(m_q)|^2 \deriv m_q \, ,
\end{equation}
where $m$ is the $\Dp\Dm \pip$ invariant mass, and $m_q$ is the $ \Dm \pip$ invariant mass. The variables $p$ and $q$ are the $\Dp$ momentum in the \myZcp rest frame and the $\Dm$ momentum  in the \mydst rest frame, respectively.
The normalization factor is chosen as
\begin{equation}
{\cal N} =
\lim_{m \rightarrow +\infty}
\int_{m_{D^{\scalebox{0.4}{\textnormal{--}}}} + m_{\pi^{\scalebox{0.4}{\textnormal{+}}}}}^{\,m - m_{D^{\scalebox{0.4}{\textnormal{+}}}}}
\frac{p\, q}{m}\, |BW(m_q)|^2\, \mathrm{d}m_q \, .
\end{equation}

such that $
    \rho_1/\rho_2 \to 1  \,\,\, \text{as} \,\,\, m \to +\infty$.

Analyticity requires $\rho_2$ to be analytic everywhere except at its branch points. By the dispersion relation, the imaginary part of $\rho_2$ is given by
\begin{equation}
   \Imag(\rho_2(s)) = \frac{(s - s_{\mathrm{th}})}{\pi} {\cal P} \int_{s_\mathrm{th}}^{\infty} \frac{\Real(\rho_2(s'))}{(s' - s_\mathrm{th})(s' - s)} \deriv s' \, ,
\end{equation}
where $\cal P$ denotes the principal value of the integral, $s$ is the invariant mass squared of the parent particle, and $s_\mathrm{th}$ is the physical threshold in the $s$-channel.
This construction ensures that the $\rho_2$ function satisfies the correct analytic properties, with its real and imaginary parts connected through the dispersion relation, thereby preserving unitarity and causality in the amplitude description~\cite{Mikhasenko2019}.

\section{Triangle diagram contributing to \myDecay decays}

\begin{figure}[htb!]
\begin{center}
\includegraphics[width=0.8\columnwidth]{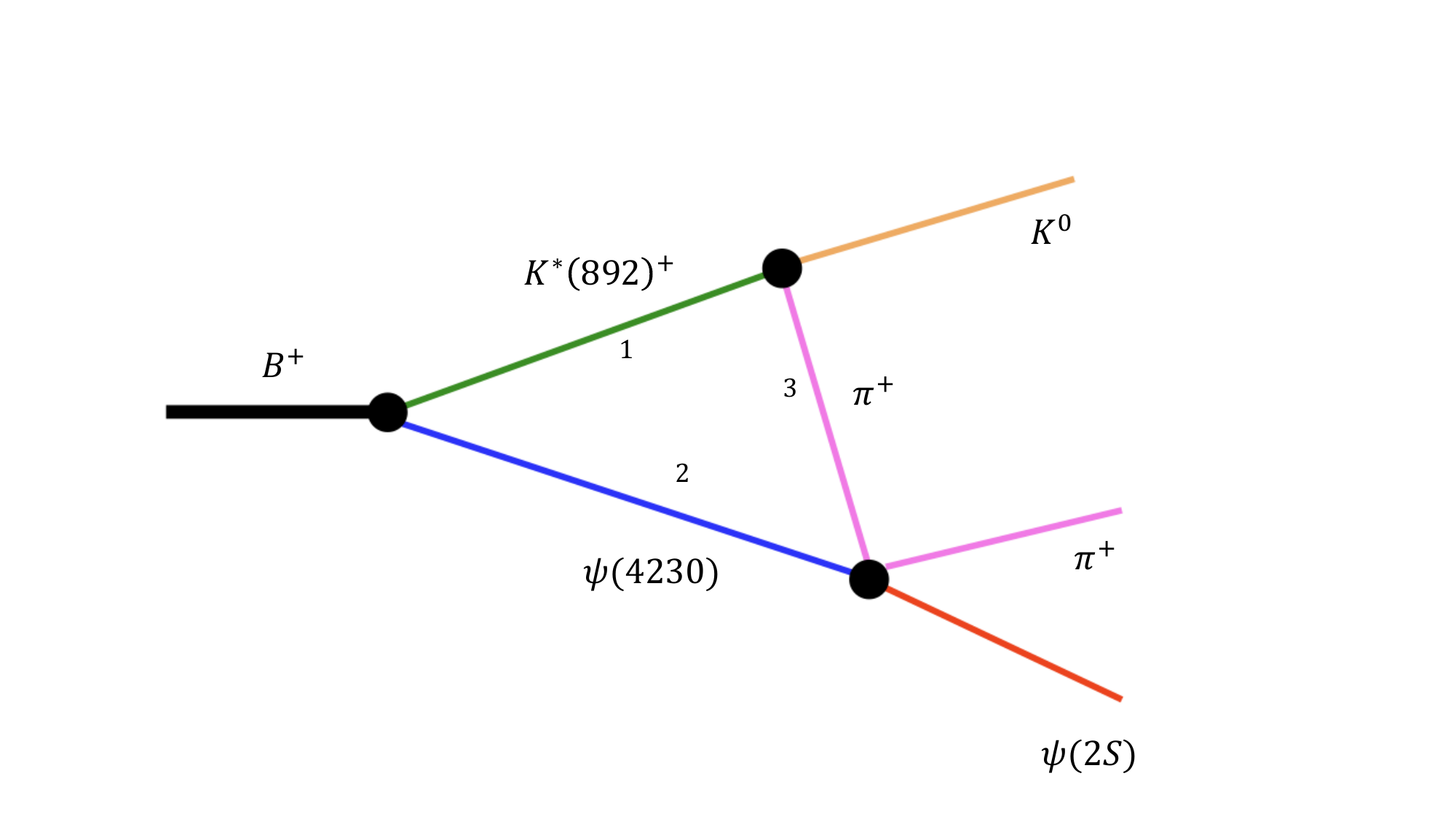}
\end{center}
\caption{The triangle diagram $K^*(892)^+\psi(4230)\pi^+$ contributing to $B^{+} \to \psi(2S) K_{\text{S}}^{0} \pi^{+}$ decays.}

\label{fig:tsdiagram}
\end{figure}

The $\Kstar(892)^+\psi(4230)\pip$ triangle diagram which contributes to \myDecay decays is shown in Fig.~\ref{fig:tsdiagram}~\cite{Nakamura:2019btl}.

\section{Angular distributions for amplitude fit}

\begin{figure}[b!]
\begin{center}
\includegraphics[width=0.32\columnwidth]{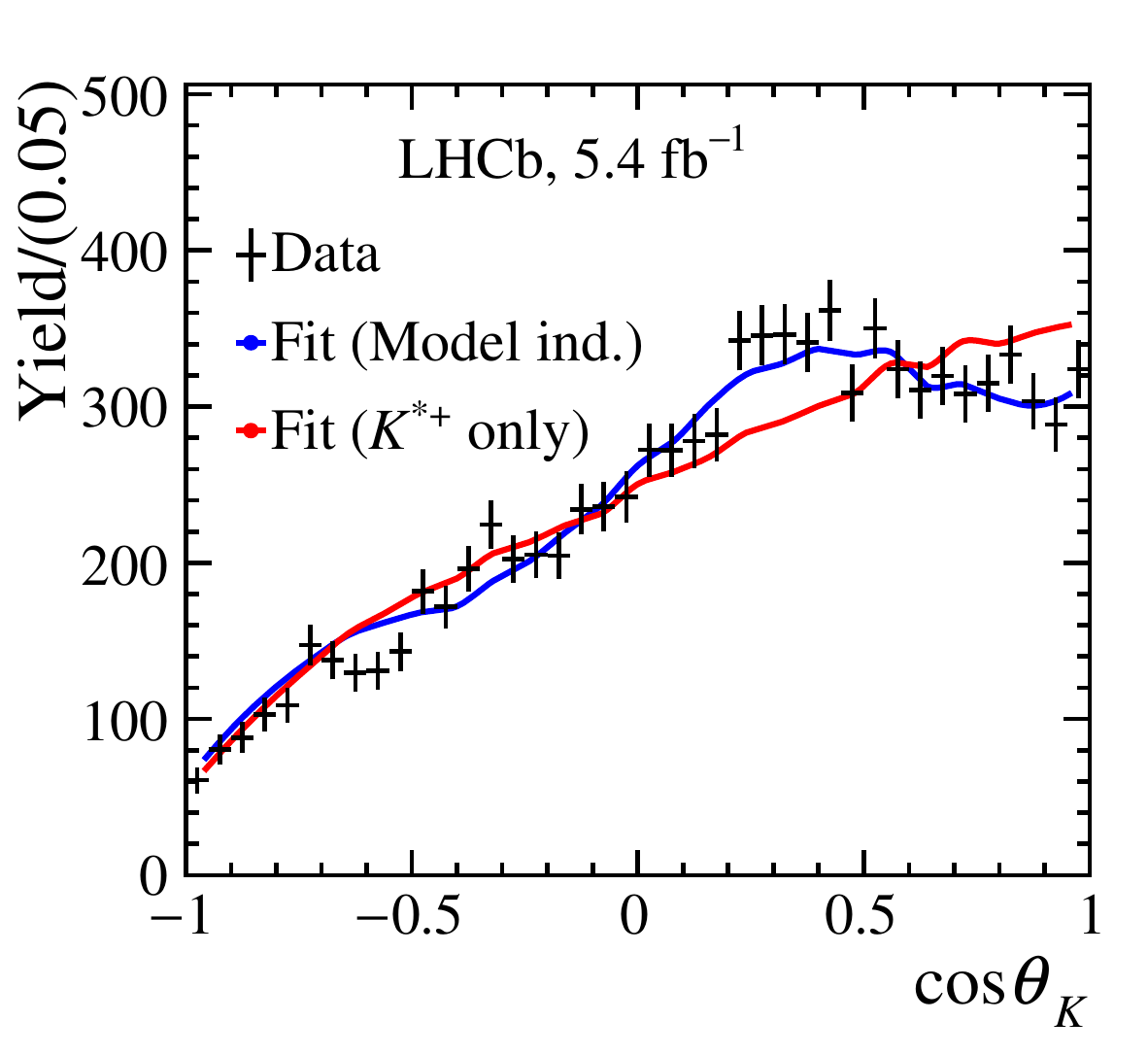}
\includegraphics[width=0.32\columnwidth]{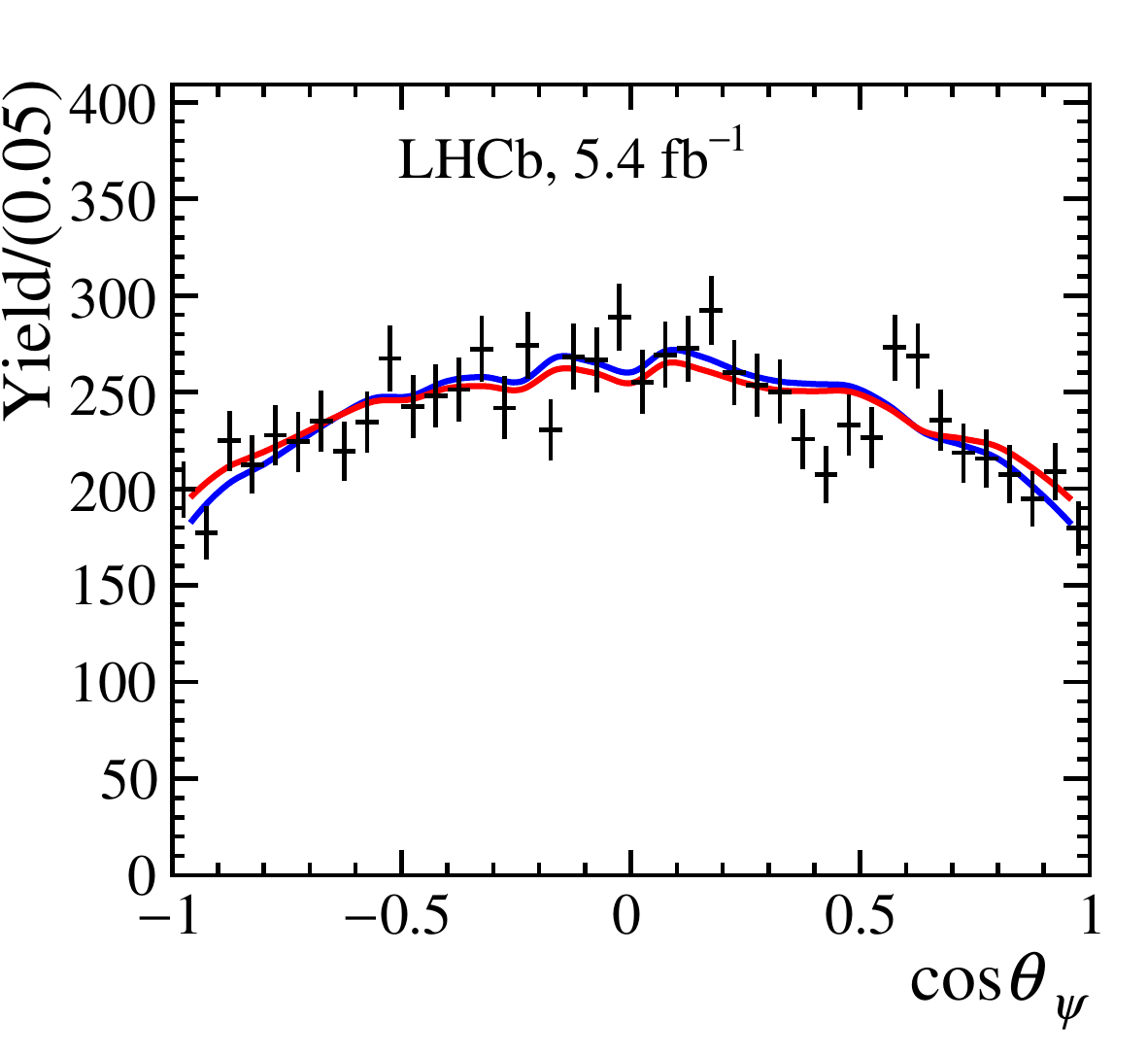}
\includegraphics[width=0.32\columnwidth]{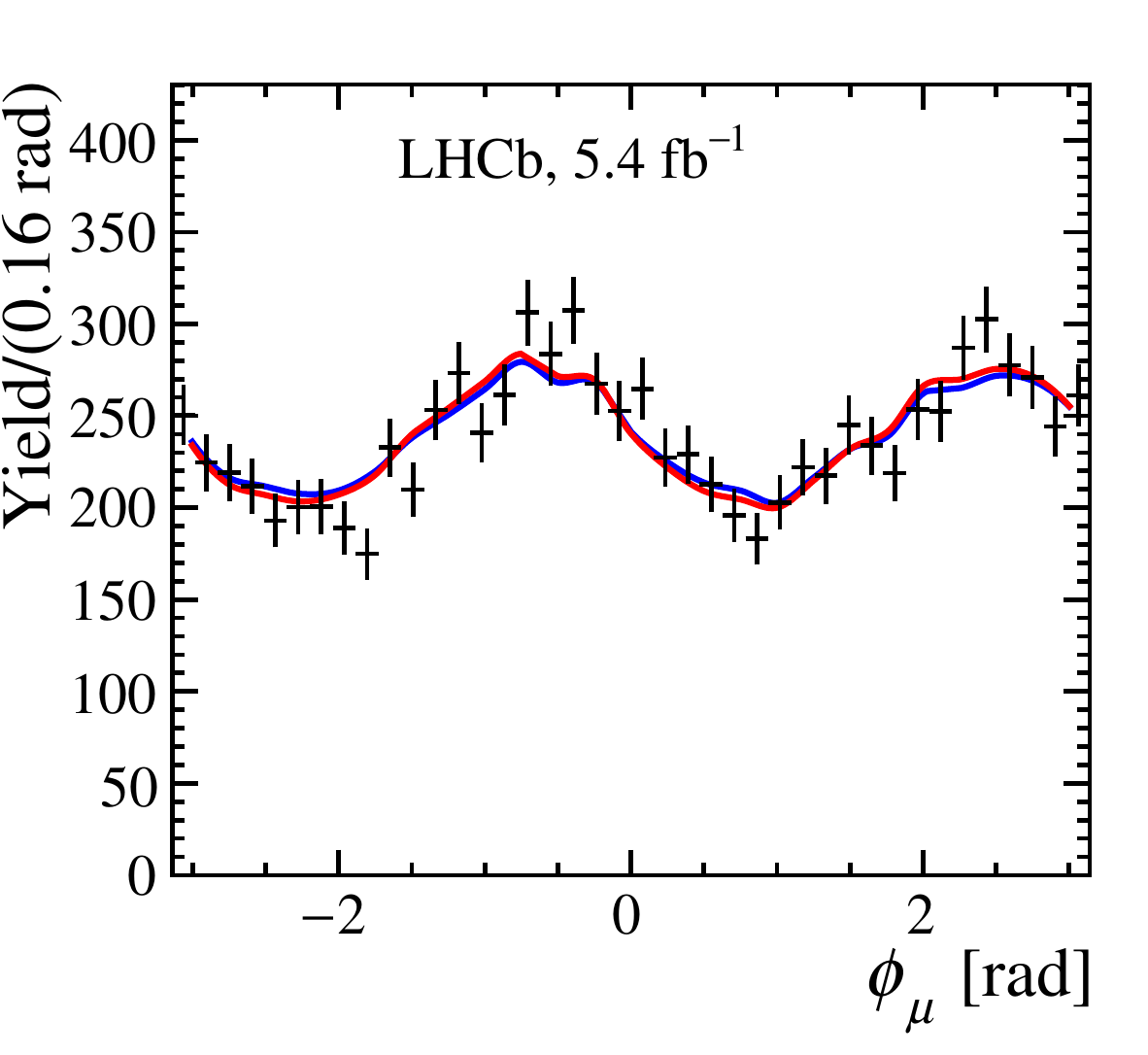}
\end{center}
\caption{Angular distributions of the (left) $\cos \theta_{K^*}$, (middle) $\cos \theta_{\psi}$ and (right) $\phi$ with the result of the fit with  (red) only $K^{*+}$ contributions and (blue) model-independent amplitude also shown. The data distributions are background-subtracted. }
\label{fig:AngleMI}
\end{figure}

Figure~\ref{fig:AngleMI} shows the angular-distribution projections of the model-independent amplitude fit and, for comparison, of the fit including only the $\Kstarp$ contributions.

\begin{figure}[b!]
\begin{center}
\includegraphics[width=0.32\columnwidth]{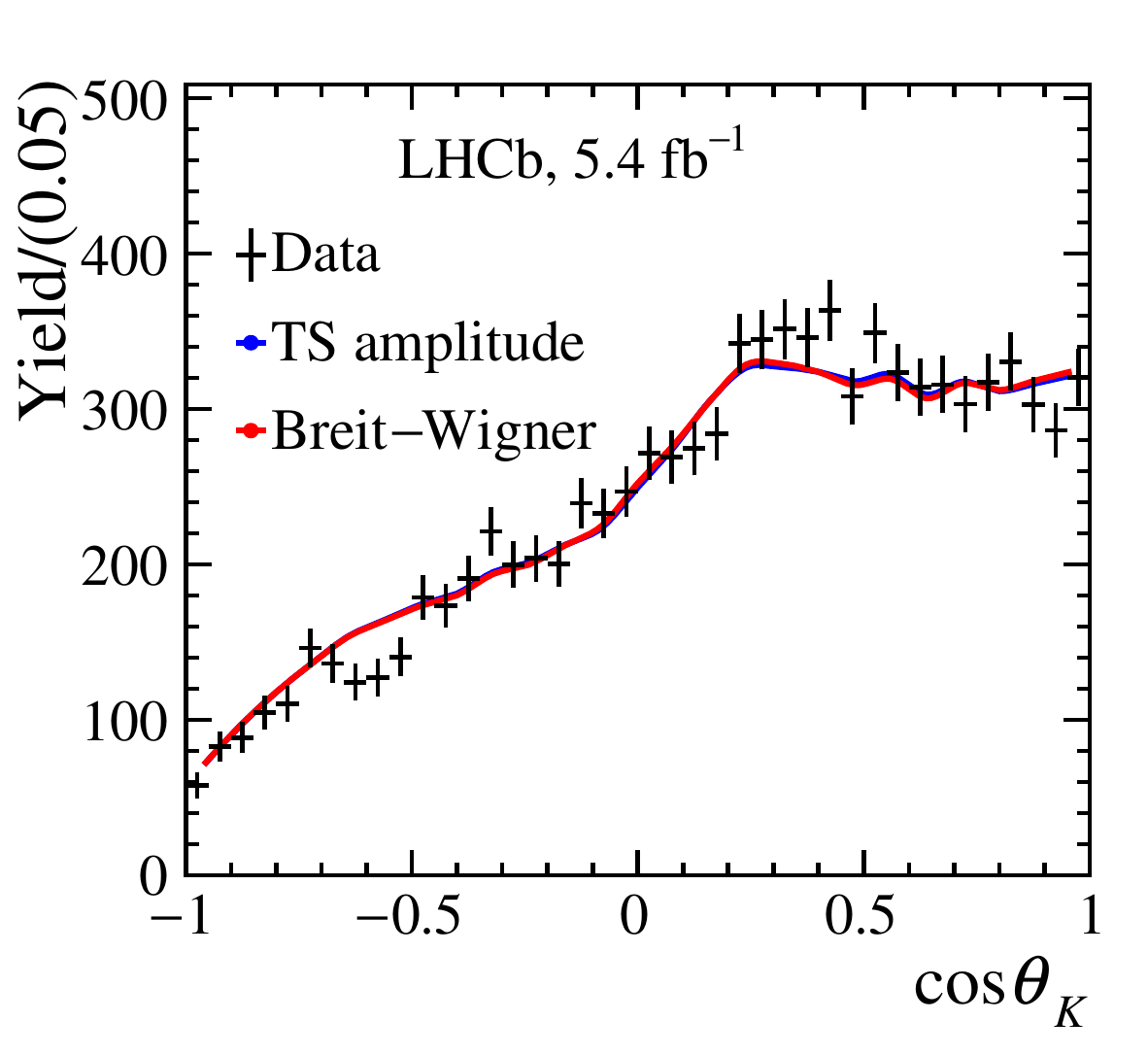}
\includegraphics[width=0.32\columnwidth]{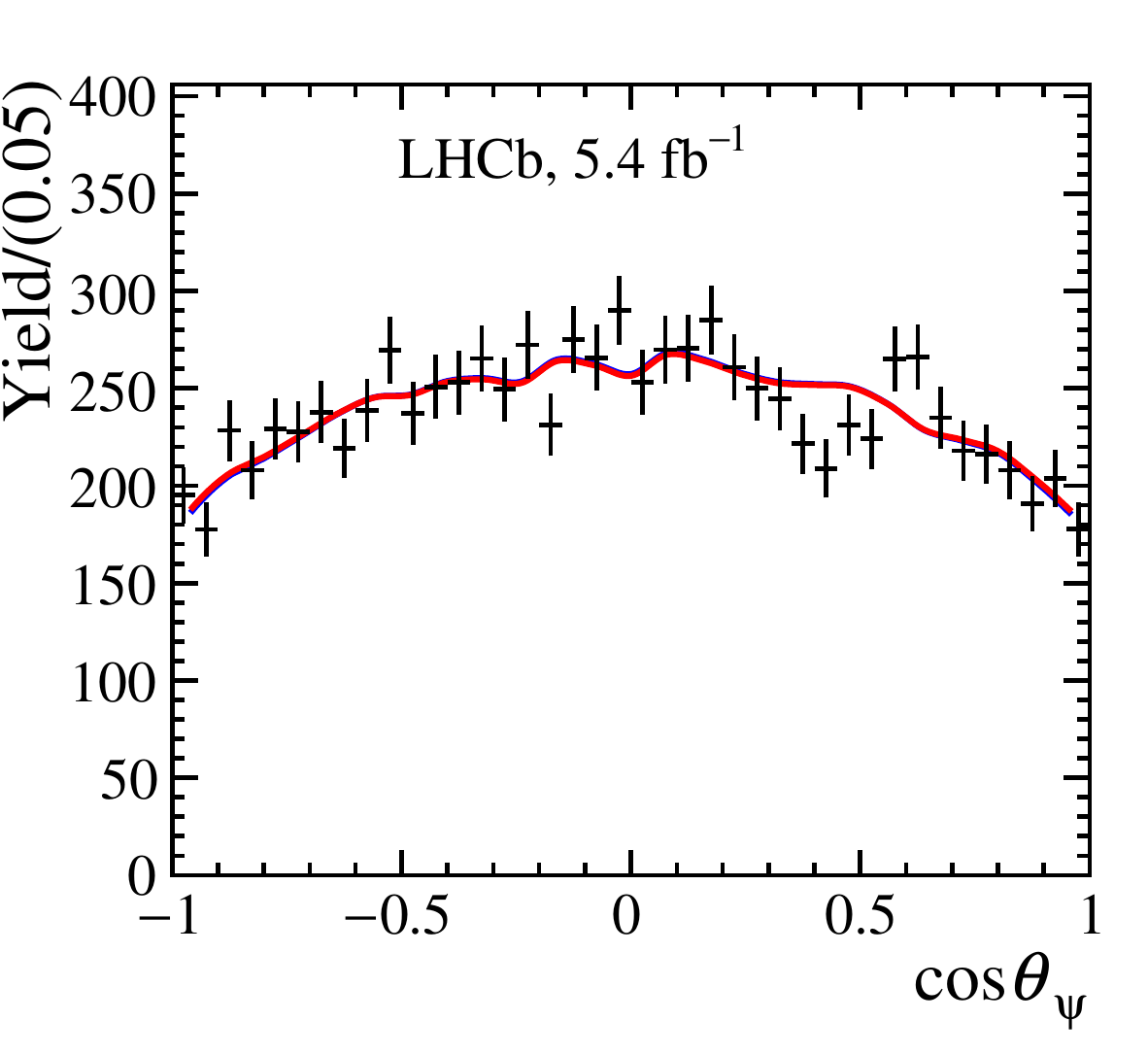}
\includegraphics[width=0.32\columnwidth]{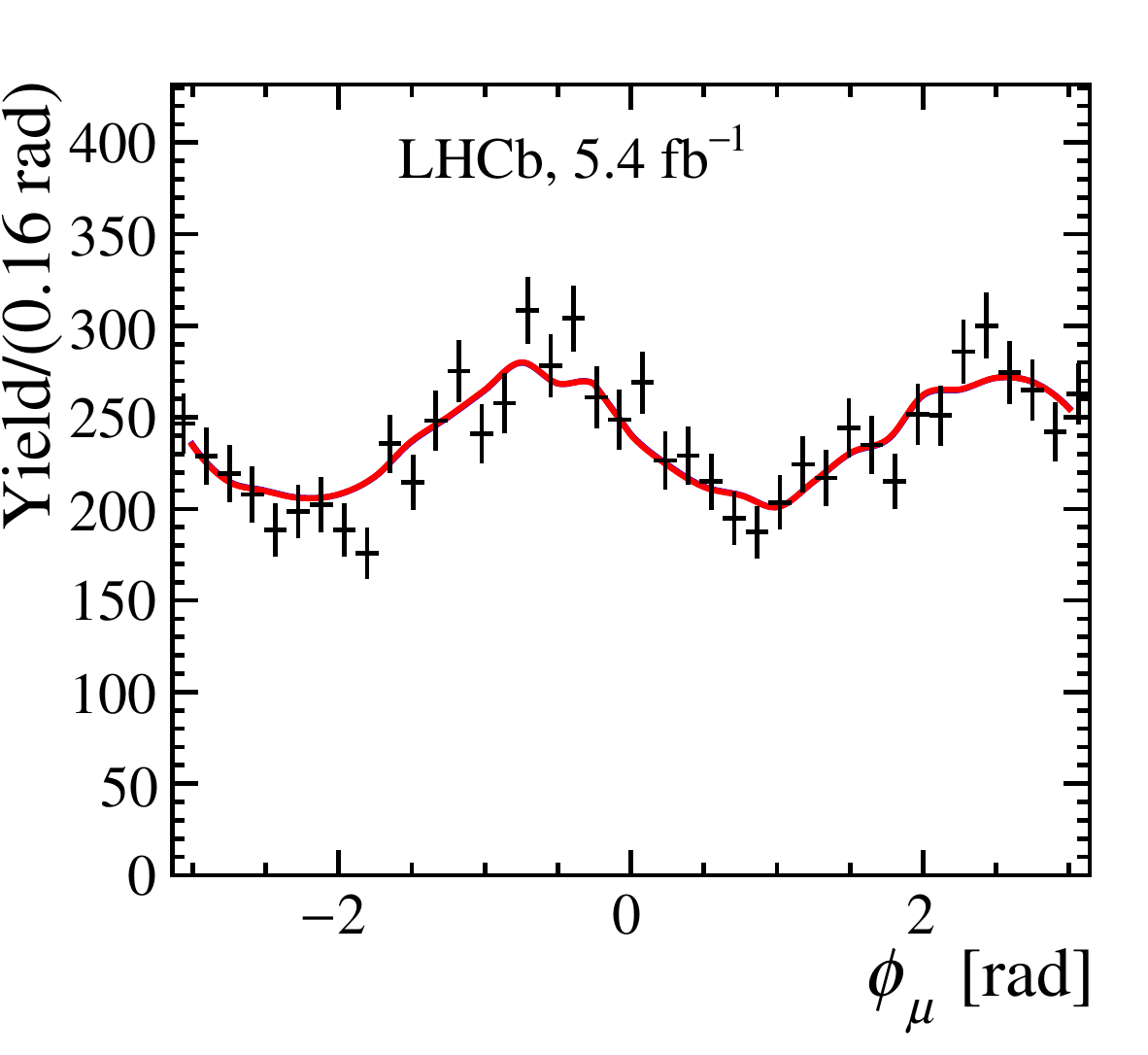}
\end{center}
\caption{Angular distributions of the (left) $\cos \theta_{K^*}$, (middle) $\cos \theta_{\psi}$ and (right) $\phi$ for data with the result of the amplitude fit with (red) the relativistic Breit--Wigner parametrization and (blue) the TS amplitude also shown. The data distributions are background-subtracted. }
\label{fig:AngleTS}
\end{figure}

Figure~\ref{fig:AngleTS} shows the angular-distribution projections of the fit including the triangle singularity (TS) amplitude, together with those of the baseline fit.

\clearpage

\section{Fit results with contributions of each amplitude components}

\begin{figure}[b!]
\begin{center}
\includegraphics[width=0.32\columnwidth]{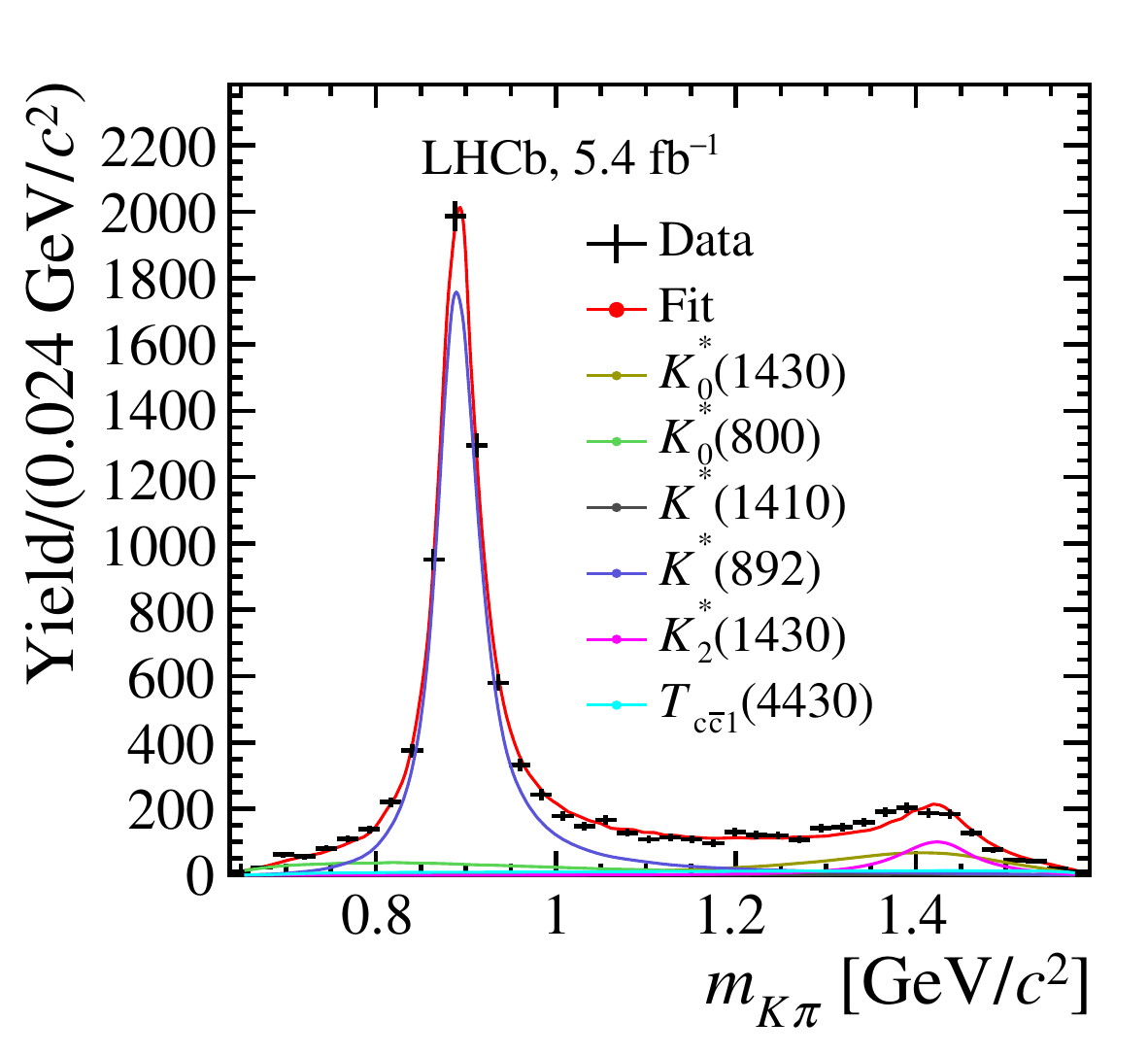}
\includegraphics[width=0.32\columnwidth]{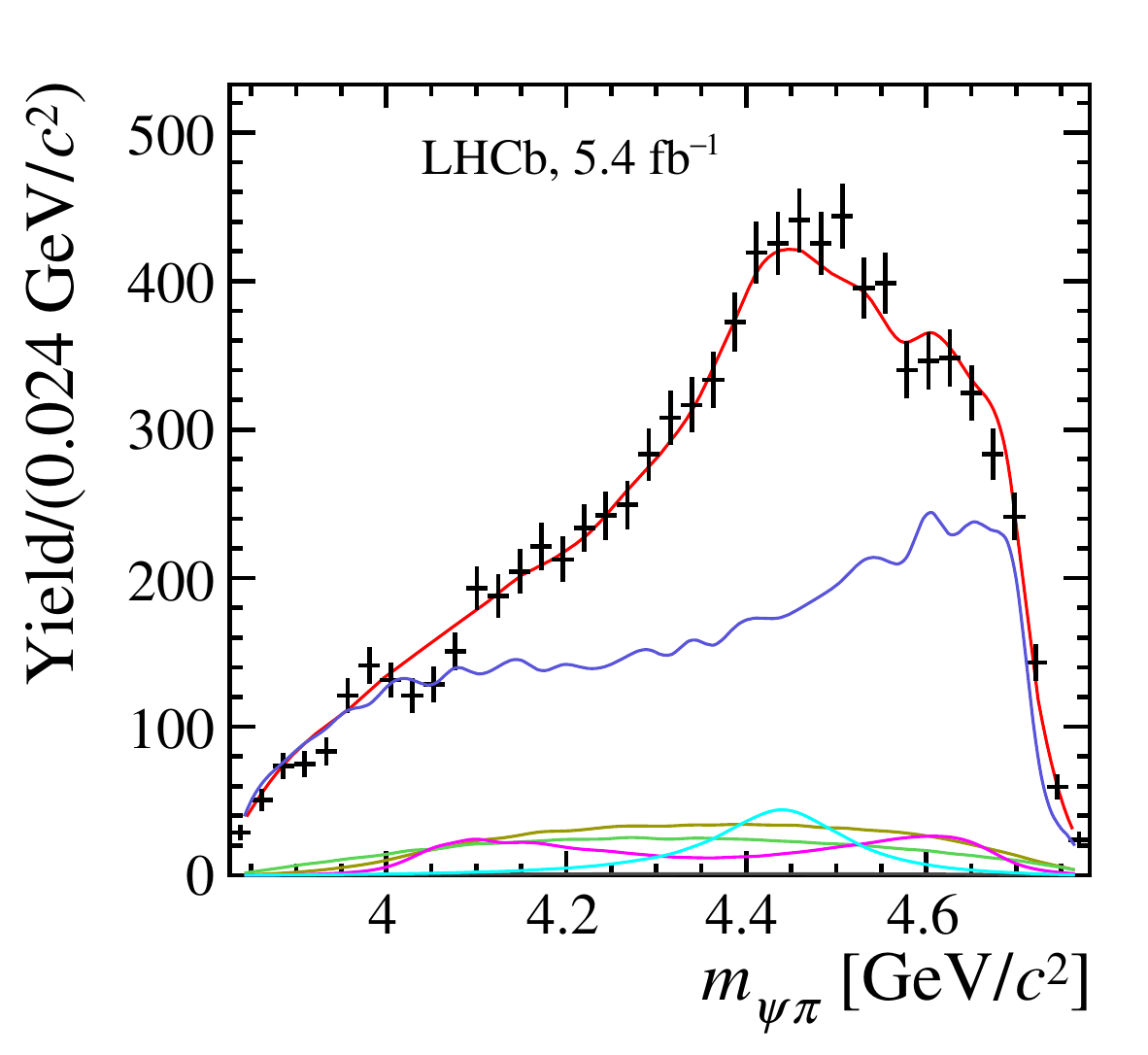}
\includegraphics[width=0.32\columnwidth]{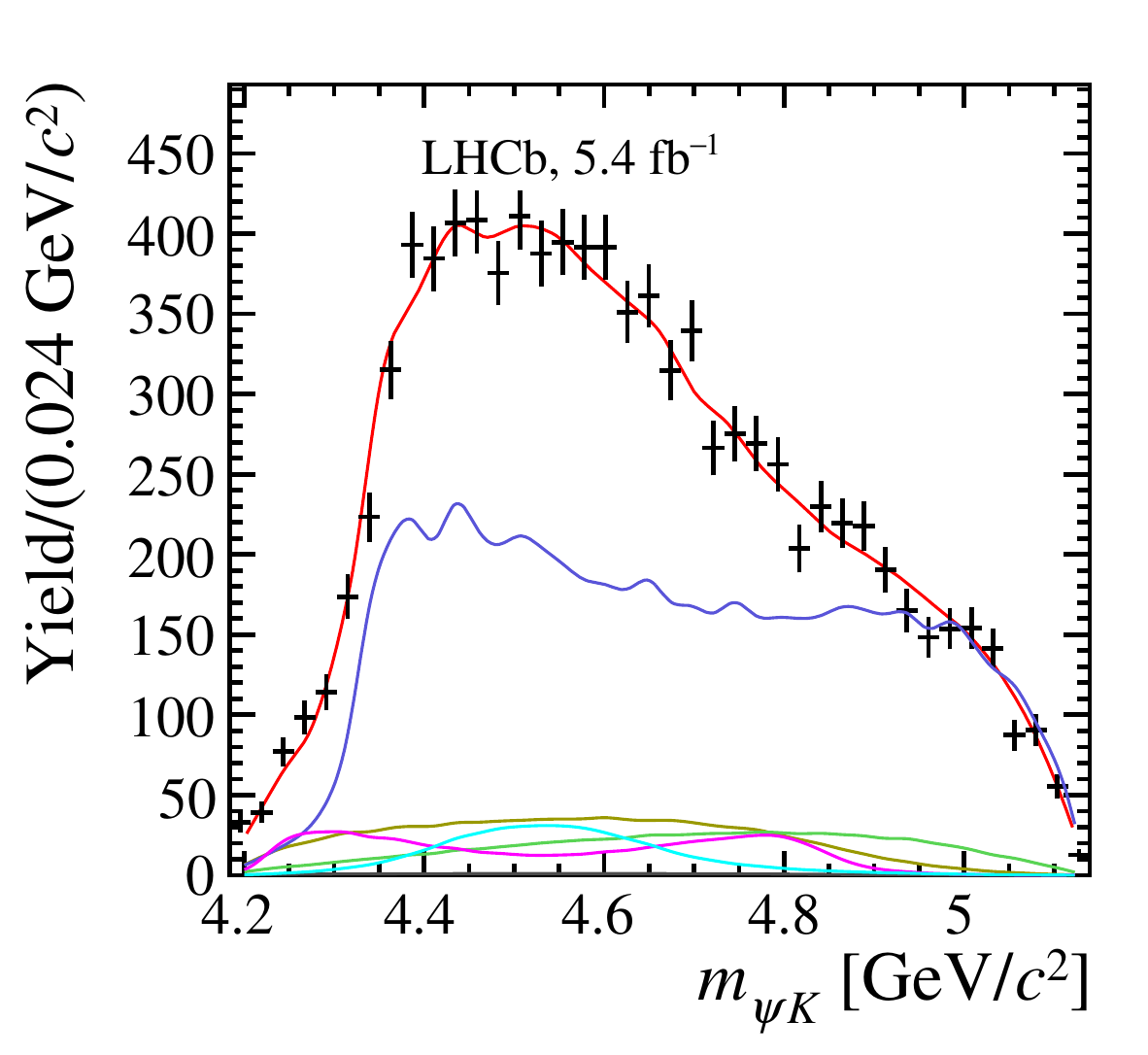}
\end{center}
\caption{Invariant-mass distributions of the (left)  $K_{\text{S}}^{0}\pi^+$ ,  (middle) $\psi(2S)\pi^+$  and (right) $\psi(2S) K_{\text{S}}^{0}$  systems for data (black dots) with the result of the amplitude fit with the relativistic Breit--Wigner parametrization (red curve), and contributions of individual amplitude components also shown. The data distributions are background-subtracted. }
\label{fig:MassBWcom}
\end{figure}

\begin{figure}[b!]
\begin{center}
\includegraphics[width=0.32\columnwidth]{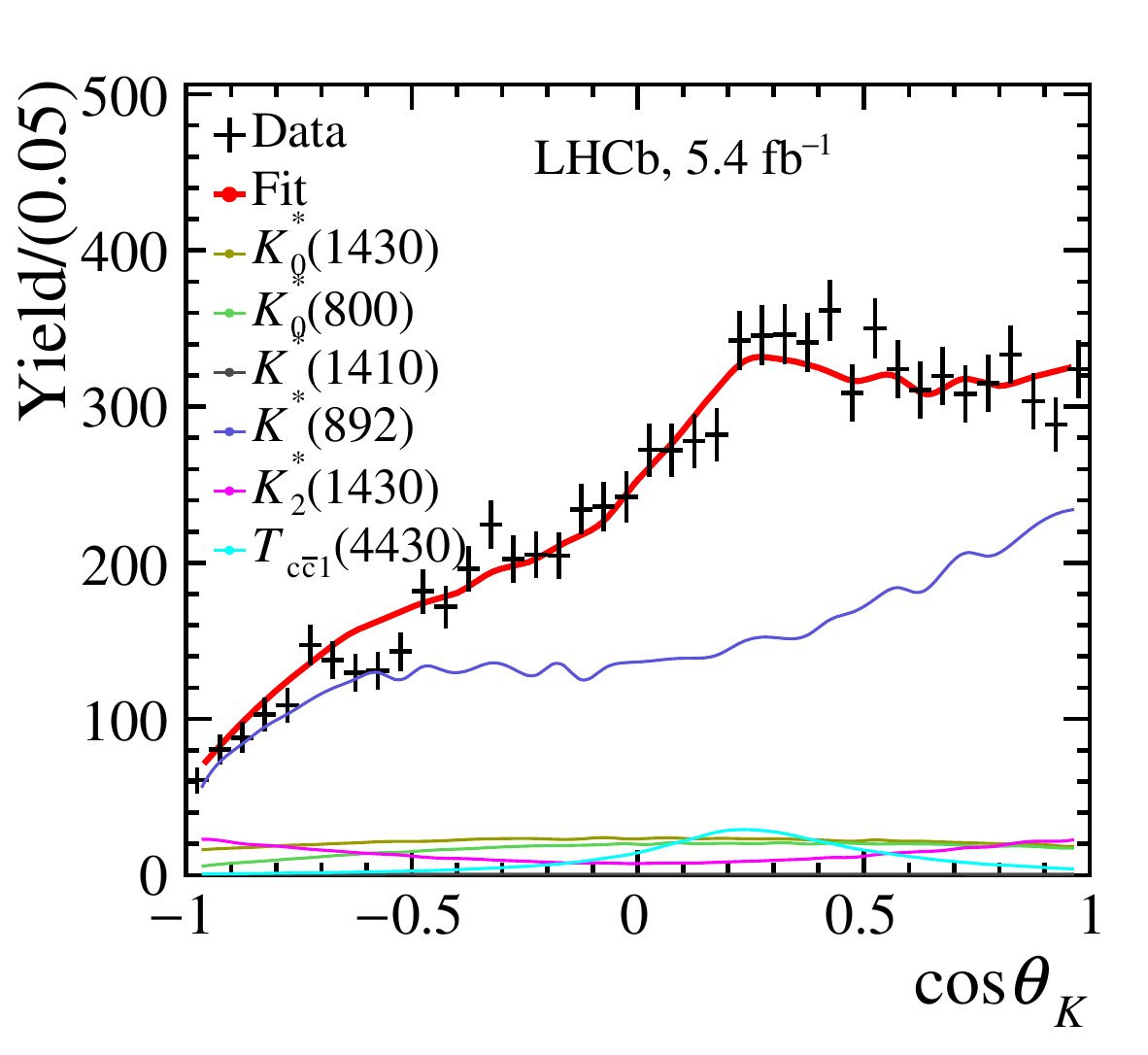}
\includegraphics[width=0.32\columnwidth]{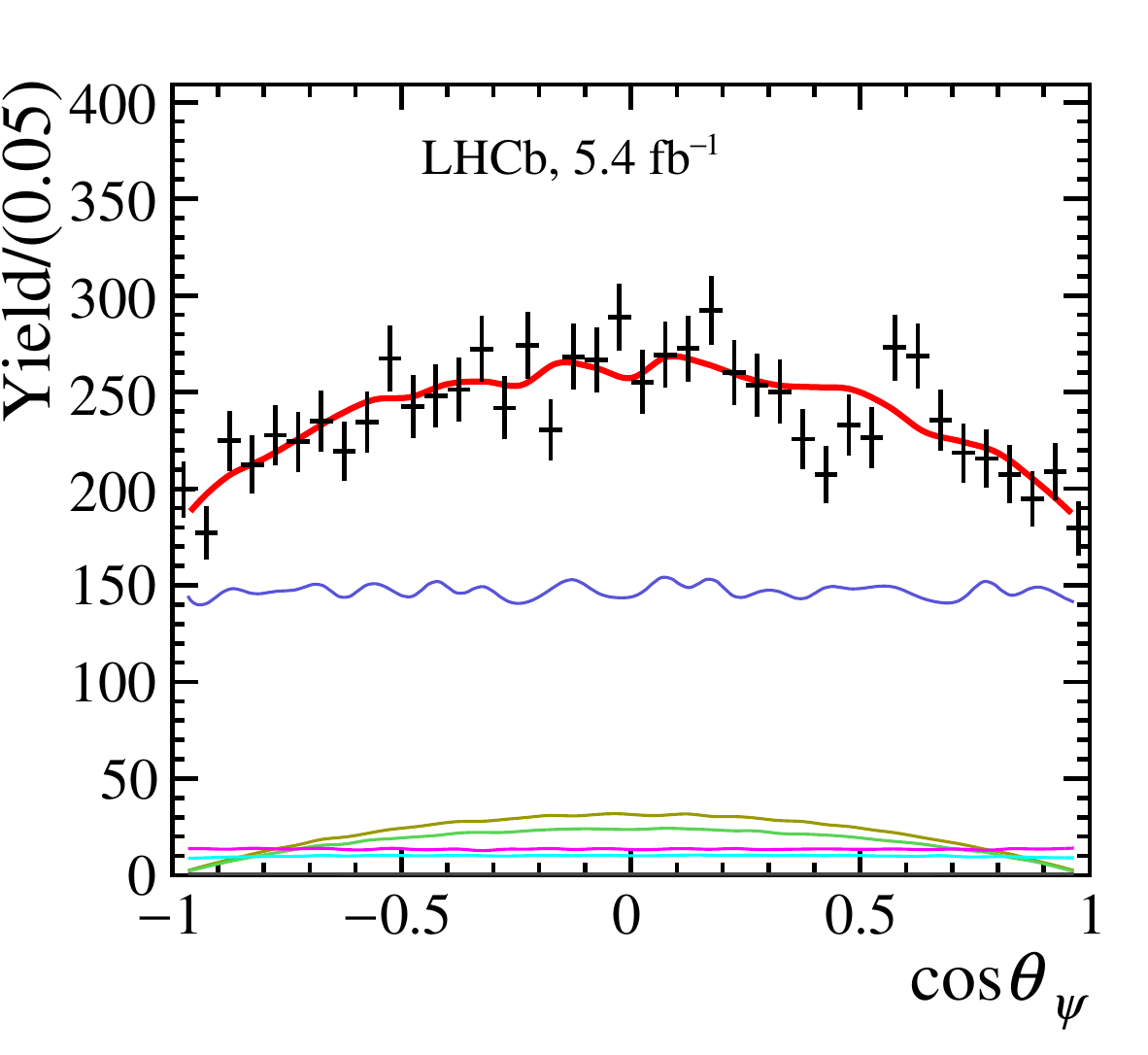}
\includegraphics[width=0.32\columnwidth]{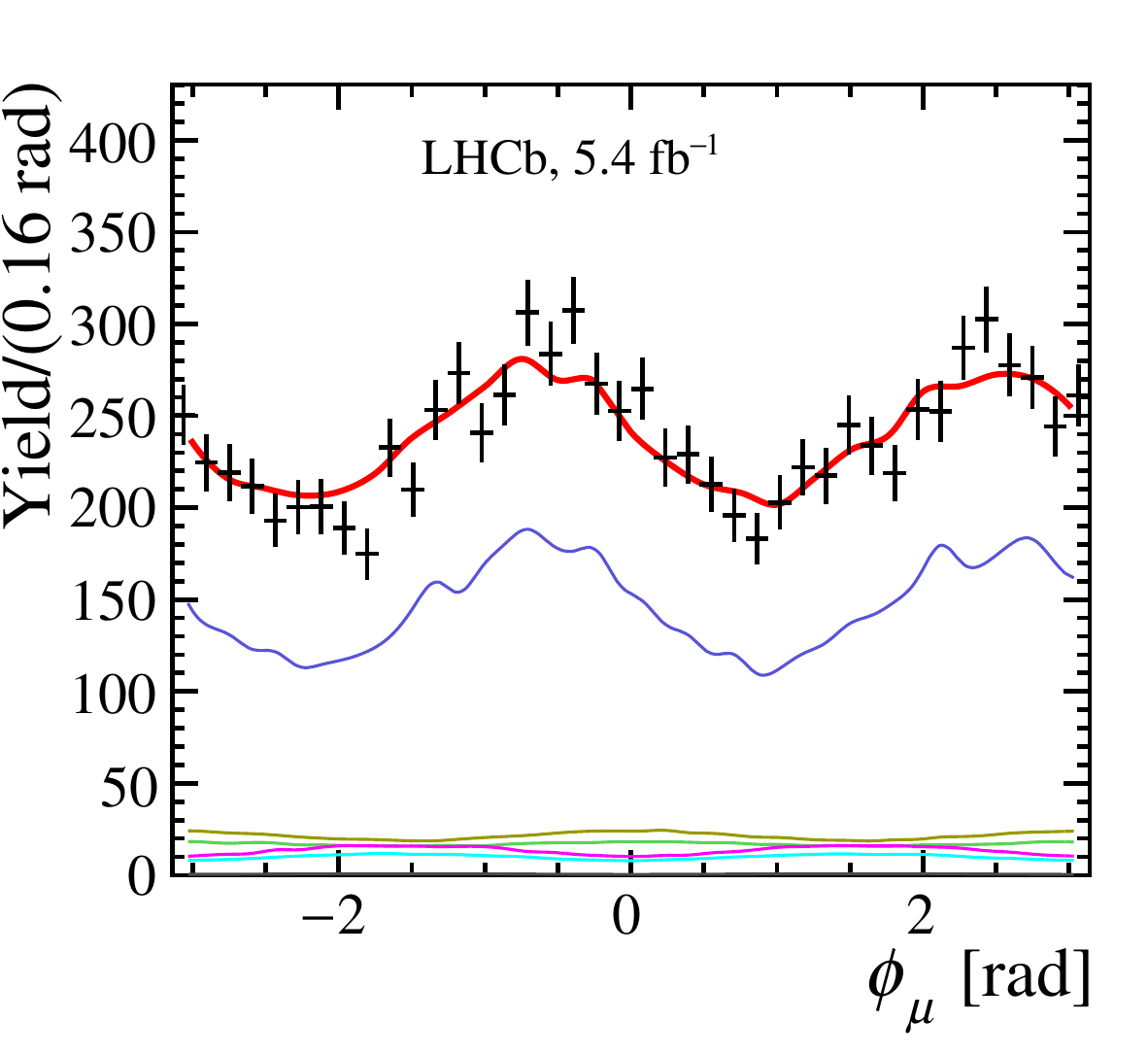}
\end{center}
\caption{Angular distributions of the (left) $\cos \theta_{K^*}$, (middle) $\cos \theta_{\psi}$ and (right) $\phi$ for data (black dots) with the result of the amplitude fit with the relativistic Breit--Wigner parametrization (red curve), and contributions of individual amplitude components also shown. The data distributions are background-subtracted. }
\label{fig:AngleBWcom}
\end{figure}

\begin{figure}[b!]
\begin{center}
\includegraphics[width=0.32\columnwidth]{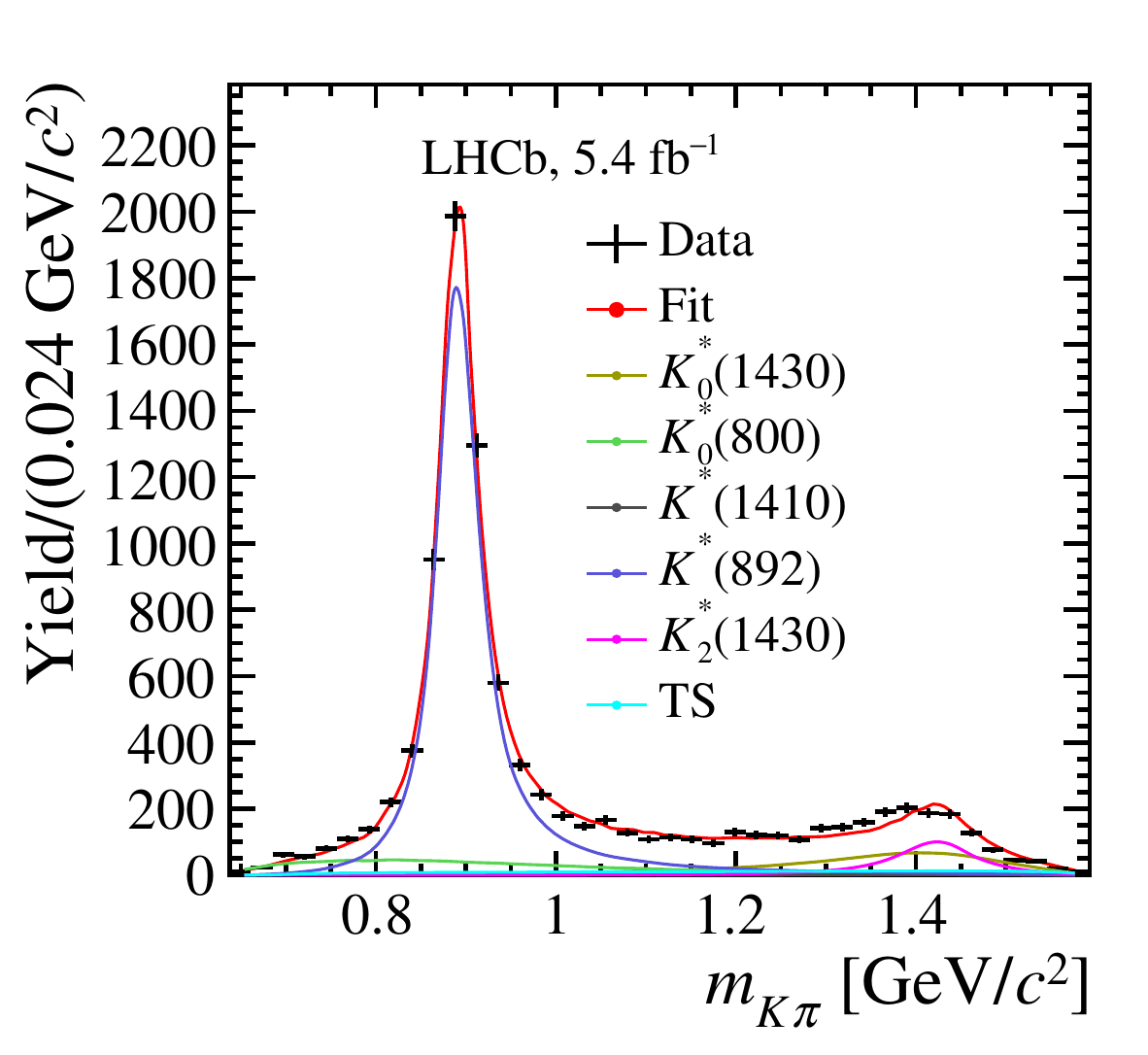}
\includegraphics[width=0.32\columnwidth]{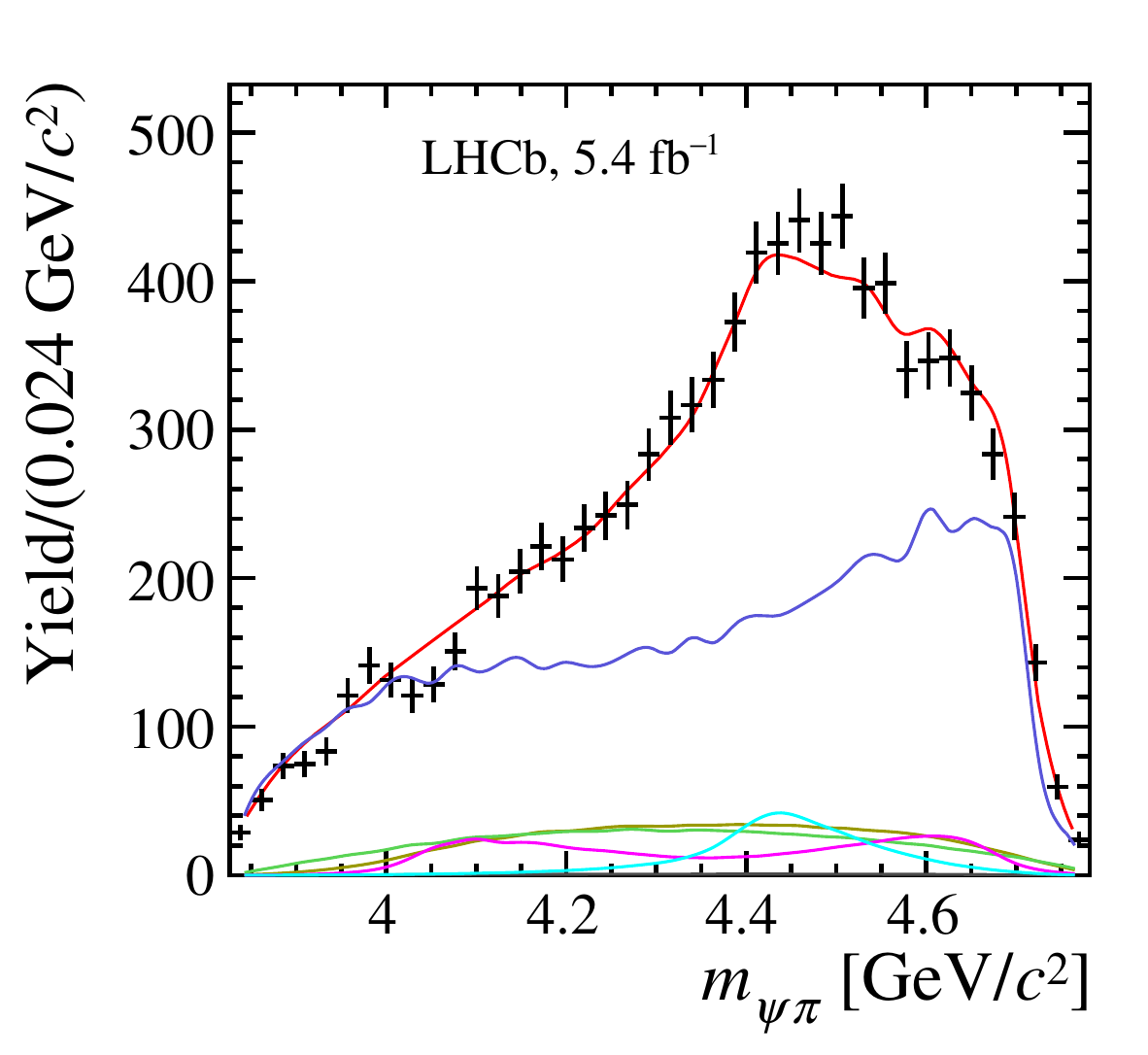}
\includegraphics[width=0.32\columnwidth]{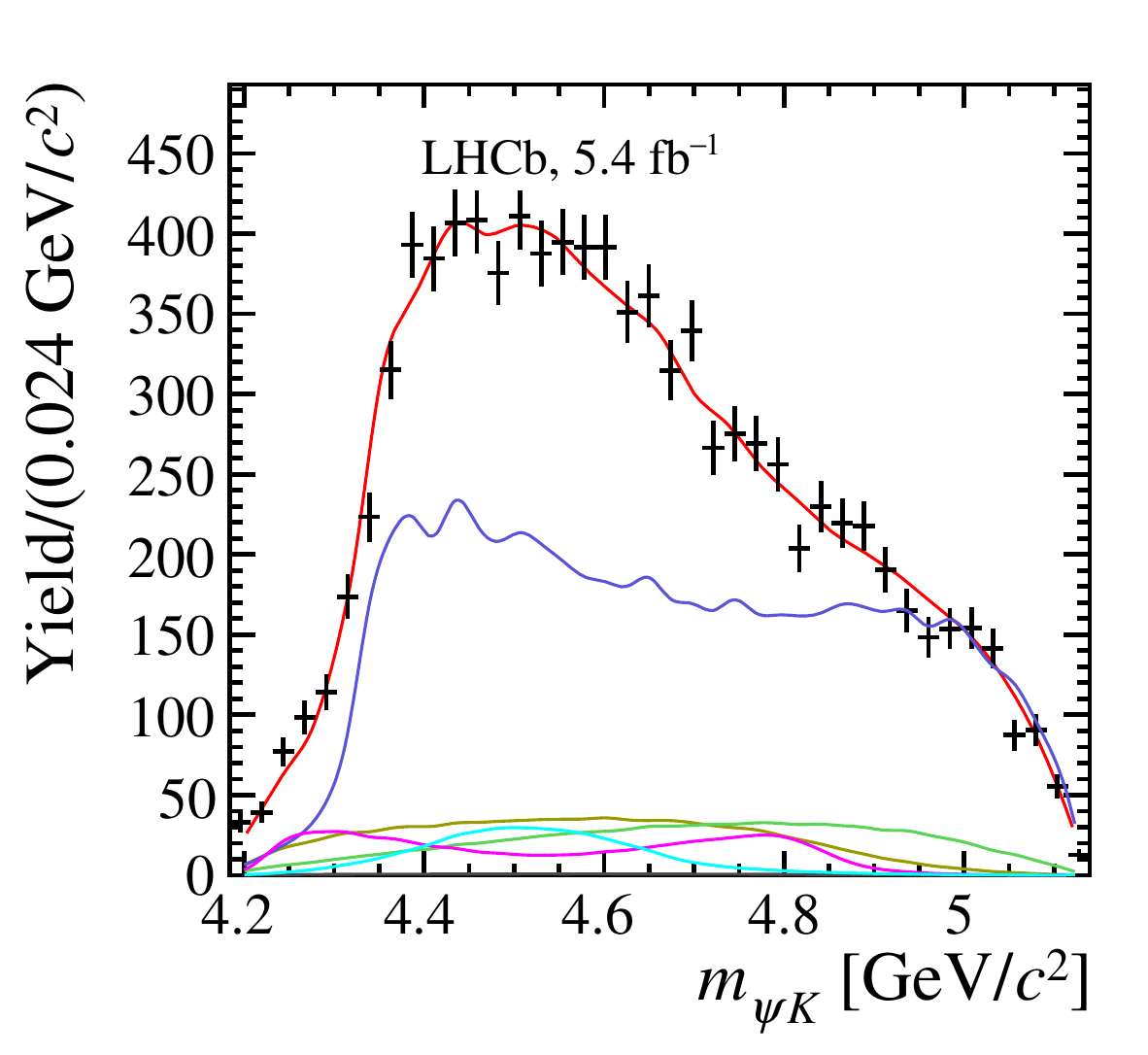}
\end{center}
\caption{Invariant-mass distributions of the  (left) $K_{\text{S}}^{0}\pi^+$ ,  (middle) $\psi(2S)\pi^+$  and (right) $\psi(2S) K_{\text{S}}^{0}$ systems for data (black dots) with the result of the amplitude fit with the TS amplitude (red curve), the contributions of amplitude components are also shown. The data distributions are background-subtracted. }
\label{fig:MassTScom}
\end{figure}

\begin{figure}[b!]
\begin{center}
\includegraphics[width=0.32\columnwidth]{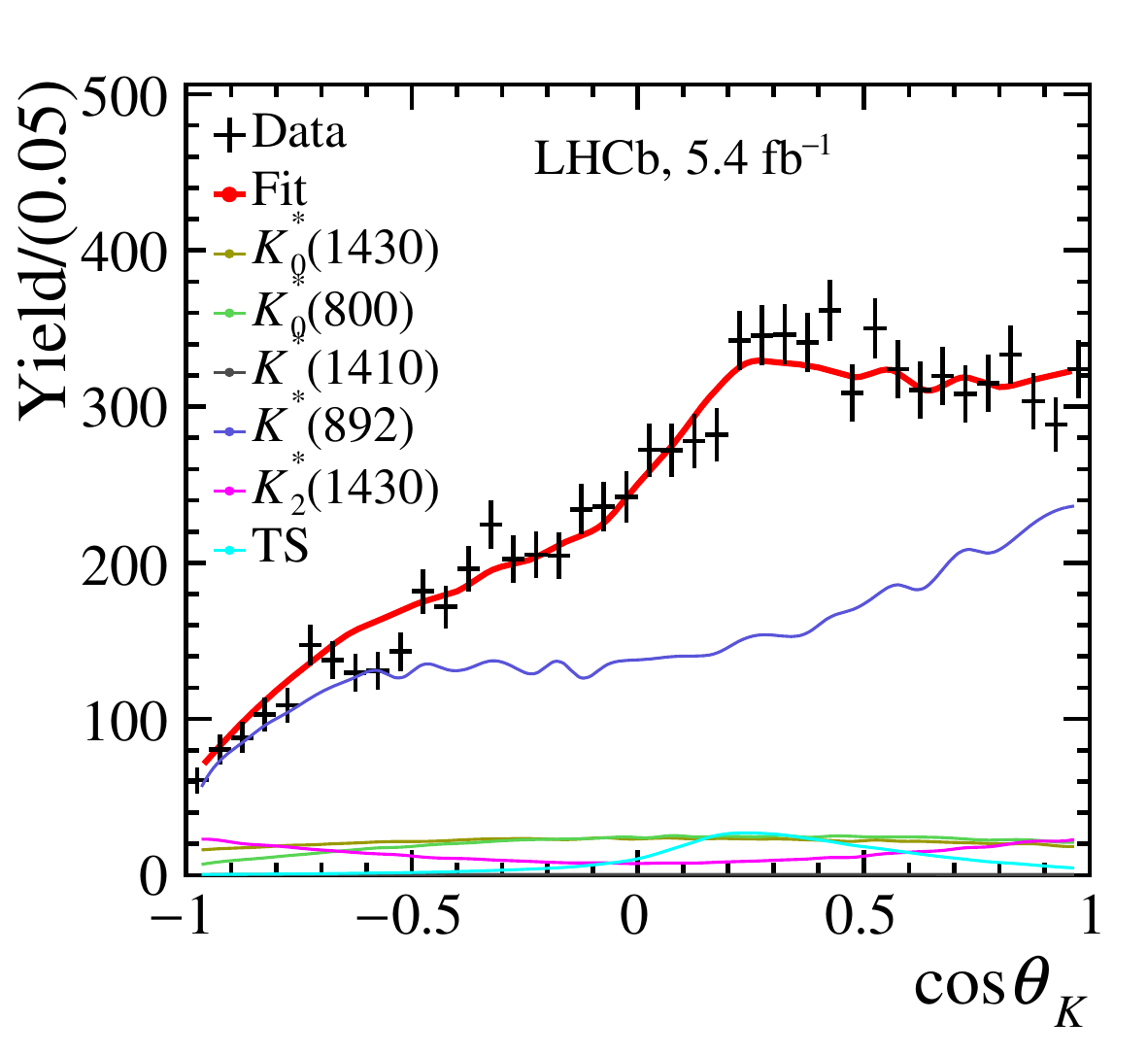}
\includegraphics[width=0.32\columnwidth]{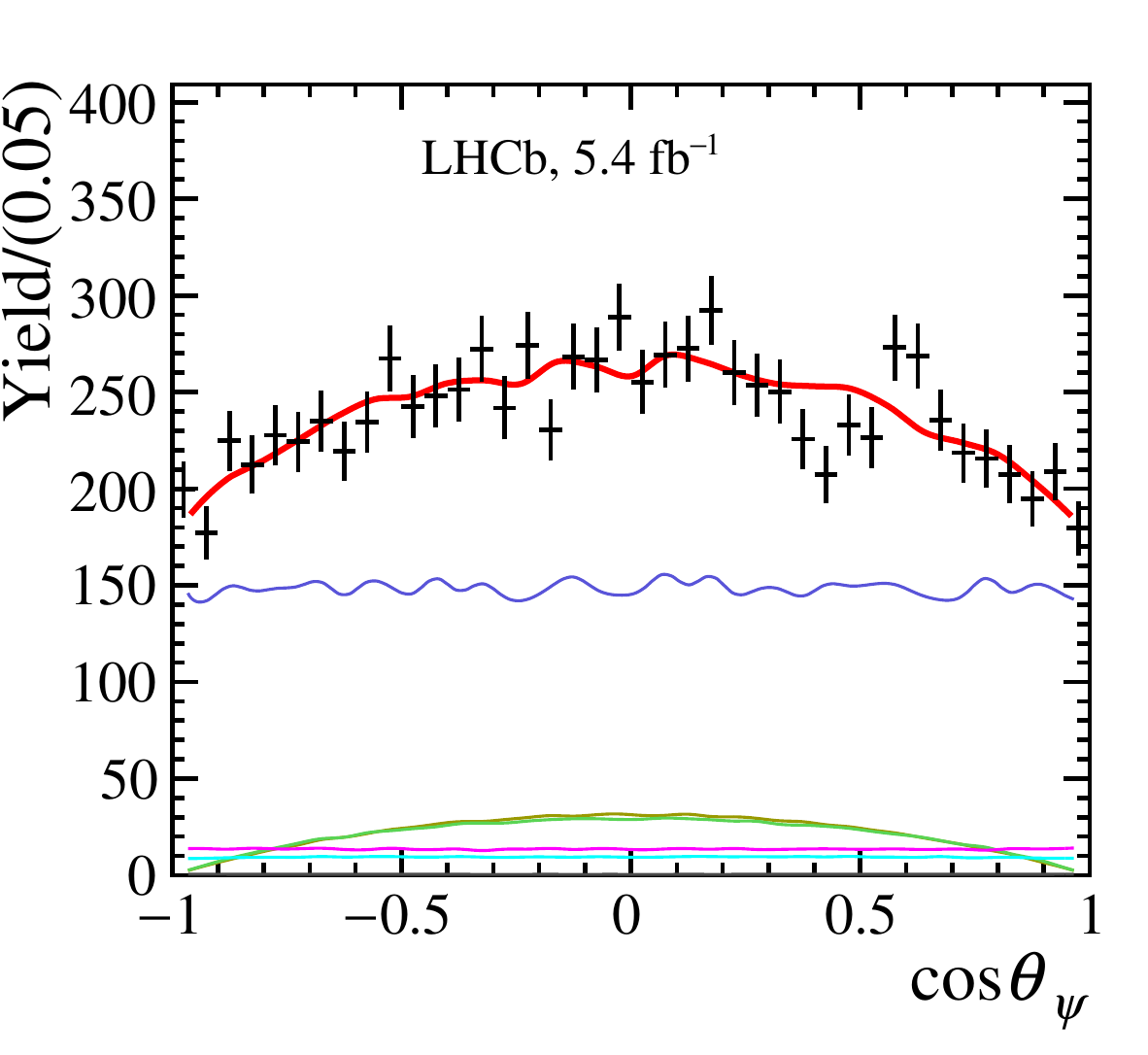}
\includegraphics[width=0.32\columnwidth]{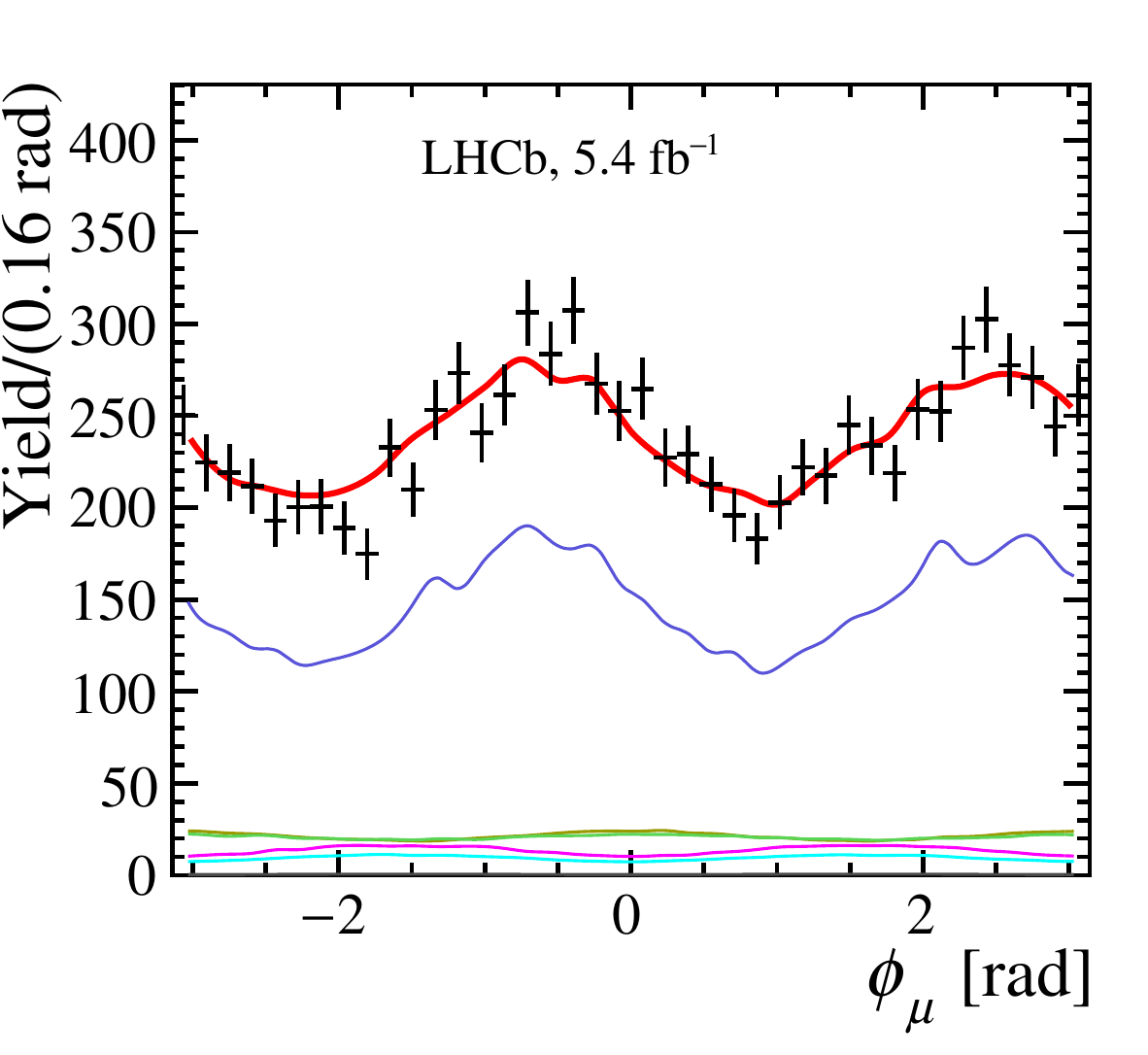}
\end{center}
\caption{Angular distributions of the (left) $\cos \theta_{K^*}$, (middle) $\cos \theta_{\psi}$ and (right) $\phi$ for data (black dots) with the result of the amplitude fit with the TS amplitude (red curve), and contributions of individual amplitude components also shown. The data distributions are background-subtracted. }
\label{fig:AngleTScom}
\end{figure}

Figure~\ref{fig:MassBWcom} shows the invariant-mass distributions of the amplitude fit with the relativistic Breit--Wigner parametrization, together with the contributions of each amplitude components.
Figure~\ref{fig:AngleBWcom} shows the angular-distribution projections of the amplitude fit with relativistic Breit--Wigner parametrization, together with the contributions of amplitude components.
Figure~\ref{fig:MassTScom} shows the invariant-mass distributions of the amplitude fit with the TS amplitude, together with the contributions of amplitude components.
Figure~\ref{fig:AngleTScom} shows the angular-distribution projections of the amplitude fit with the TS amplitude, together with the contributions of amplitude components.

\clearpage

\section{Numeric results for amplitude fit}

The amplitude fit is performed using $LS$ (orbital angular momentum and spin coupling) bases instead of helicity bases~\cite{LHCB-PAPER-2015-029}. The amplitude fit results with relativistic Breit--Wigner parametrization are summarized in Table~\ref{tab:amfitsummary}.

\begin{table}[!hbtp]
    \centering
    \caption{Results of the amplitude fit with the relativistic Breit--Wigner parametrization. For each parameter, the first uncertainty is statistical, and the second is systematic. The \Real, \Imag, FF represent the real and imaginary part of the helicity couplings and the fit fraction, respectively, while $L$ represents the orbital angular momentum between the $\psitwos$ and the $K^*(892)^+$ mesons.}
    \label{tab:amfitsummary}
    \begin{tabular}{ccc}
        \hline\hline
        Decay channel   & Parameter  &Fit result\\
        \hline
        \multirow{3}{*}{$\decay{\Bp}{\psitwos K^*_0(700)^+}$} &\Real &$-0.10\pm0.03^{+0.09}_{-0.06} $ \\
            &\Imag & $-0.30\pm0.02^{+0.09}_{-0.02}$  \\
            &FF &$0.074\pm0.016^{+0.037}_{-0.018} $\\
        \hline
        \multirow{3}{*}{$\decay{\Bp}{\psitwos K^*_0(1430)^+}$} &\Real      &$0.09\pm0.03^{+0.09}_{-0.06}$\\
        &\Imag   & $0.27\pm0.03^{+0.01}_{-0.09}$\\
        &FF   &$0.085\pm0.018^{+0.001}_{-0.046}$\\
        \hline
        \multirow{9}{*}{$\decay{\Bp}{\psitwos K^*(892)^+}$} &\Real($L=0$) &1 (fixed) \\
            &\Imag ($L=0$) &0 (fixed)  \\
            &\Real($L=1$)&$-0.51\pm0.03^{+0.04}_{-0.01}$\\
            &\Imag ($L=1$) &$0.41\pm0.03^{+0.03}_{-0.02}$\\
            &\Real($L=2$) &$-0.41\pm0.02^{+0.02}_{-0.01}$\\
            &\Imag ($L=2$) &$0.37\pm0.04^{+0.01}_{-0.05}$\\
            &FF       &$0.658\pm0.013^{+0.08}_{-0.22}$\\
            &Mass     &$891.7\pm0.2^{+0.2}_{-0.0}\mevcc$\\
            &Width    &$49.6\pm0.6^{+0.2}_{-0.3}\mev$\\
        \hline
            \multirow{3}{*}{$\decay{\Bp}{\psitwos K^*(1410)^+}$}  
            &\Real&$-0.08\pm0.02^{+0.14}_{-0.05}$\\
            &\Imag &$-0.02\pm0.03^{+0.18}_{-0.10}$\\
            &FF &$0.002\pm0.001^{+0.013}_{-0.001}$\\
                \hline
            \multirow{3}{*}{$\decay{\Bp}{\psitwos K^*_2(1430)^+}$} &\Real &$0.48\pm0.02^{+0.01}_{-0.07}$\\
                &\Imag &$0.08\pm0.05^{+0.01}_{-0.15}$\\
                &FF &$0.053\pm0.005^{+0.05}_{-0.07}$\\
            \hline
            \multirow{5}{*}{$\decay{\Bp}{\myZcp \KS}$}    
                &\Real  &$-0.255\pm0.03$\\
                &\Imag   &$0.144\pm0.066$\\
                &FF   &$0.037\pm0.006^{+0.033}_{-0.007}$\\
                &Mass &$4452.0\pm16.3^{+23.1}_{-32.6} \mevcc$\\
                &Width&$173.9\pm18.6^{+75.1}_{-20.0} \mev$\\
            \hline
            Total FF  &   &$0.909\pm0.045^{+0.015}_{-0.050}$ \\
        \hline\hline
    
    \end{tabular}
\end{table}

\clearpage
The amplitude fit results with triangle singularity amplitude are summarized in Table~\ref{tab:amfitsummary_TS}.

\begin{table}[!hbtp]
    \centering
    \caption{Results of the amplitude fit with the triangle singularity amplitude. For each parameter, the first uncertainty is statistical, and the second is systematic. The \Real, \Imag, FF represent the real and imaginary part of the helicity couplings and the fit fraction, respectively, while $L$ represents the orbital angular momentum between the $\psitwos$ and  $K^*(892)^+$ mesons.}
    \label{tab:amfitsummary_TS}
    \begin{tabular}{ccc}
        \hline\hline
        Decay channel   & Parameter  &Fit result\\
        \hline
        \multirow{3}{*}{$\decay{\Bp}{\psitwos K^*_0(700)^+}$} 
            &\Real&$-0.11\pm0.02^{+0.02}_{-0.05} $ \\
            &\Imag & $-0.33\pm0.02^{+0.04}_{-0.01}$  \\
            &FF &$0.087\pm0.015^{+0.015}_{-0.016} $\\
        \hline
        \multirow{3}{*}{$\decay{\Bp}{\psitwos K^*_0(1430)^+}$} 
        &\Real&$0.07\pm0.03^{+0.06}_{-0.07}$\\
        &\Imag   & $0.27\pm0.02^{+0.00}_{-0.09}$\\
        &FF   &$0.089\pm0.027^{+0.011}_{-0.040}$\\
        \hline
        \multirow{9}{*}{$\decay{\Bp}{\psitwos K^*(892)^+}$} 
        &\Real($L=0$) &1 (fixed) \\
            &\Imag ($L=0$) &0 (fixed)  \\
            &\Real($L=1$)&$-0.50\pm0.03^{+0.01}_{-0.03}$\\
            &\Imag ($L=1$) &$0.41\pm0.03^{+0.02}_{-0.01}$\\
            &\Real($L=2$) &$-0.41\pm0.02^{+0.03}_{-0.01}$\\
            &\Imag ($L=2$) &$0.38\pm0.04^{+0.01}_{-0.06}$\\
            &FF       &$0.622\pm0.018^{+0.000}_{-0.019}$\\
            &Mass     &$891.7\pm0.2^{+0.2}_{-0.0}\mevcc$\\
            &Width    &$49.6\pm0.6^{+0.1}_{-0.2}\mev$\\
        \hline
            \multirow{3}{*}{$\decay{\Bp}{\psitwos K^*(1410)^+}$}  
            &\Real&$-0.07\pm0.02^{+0.10}_{-0.06}$\\
            &\Imag &$0.00\pm0.03^{+0.14}_{-0.01}$\\
            &FF &$0.002\pm0.003^{+0.012}_{-0.001}$\\
                \hline
            \multirow{3}{*}{$\decay{\Bp}{\psitwos K^*_2(1430)^+}$} 
                &\Real &$0.47\pm0.02^{+0.01}_{-0.09}$\\
                &\Imag &$0.10\pm0.05^{+0.01}_{-0.25}$\\
                &FF &$0.056\pm0.004^{+0.001}_{-0.005}$\\
            \hline
            \multirow{3}{*}{$\decay{\Bp}{\psi(4230)\pi^+ \KS}$}    
                &\Real  &$2.84\pm0.53^{+0.15}_{-2.54}$\\
                &\Imag   &$-4.97\pm0.46^{+0.13}_{-1.39}$\\
                &FF   &$0.039\pm0.007^{+0.027}_{-0.001}$\\
            \hline
            Total FF  &   &$0.896\pm0.064^{+0.032}_{-0.045}$ \\
        \hline\hline

    \end{tabular}
\end{table}

\clearpage

\section{Amplitude fit with alternative triangle singularity amplitude}
\begin{figure}[b!]
\begin{center}
\includegraphics[width=0.9\columnwidth]{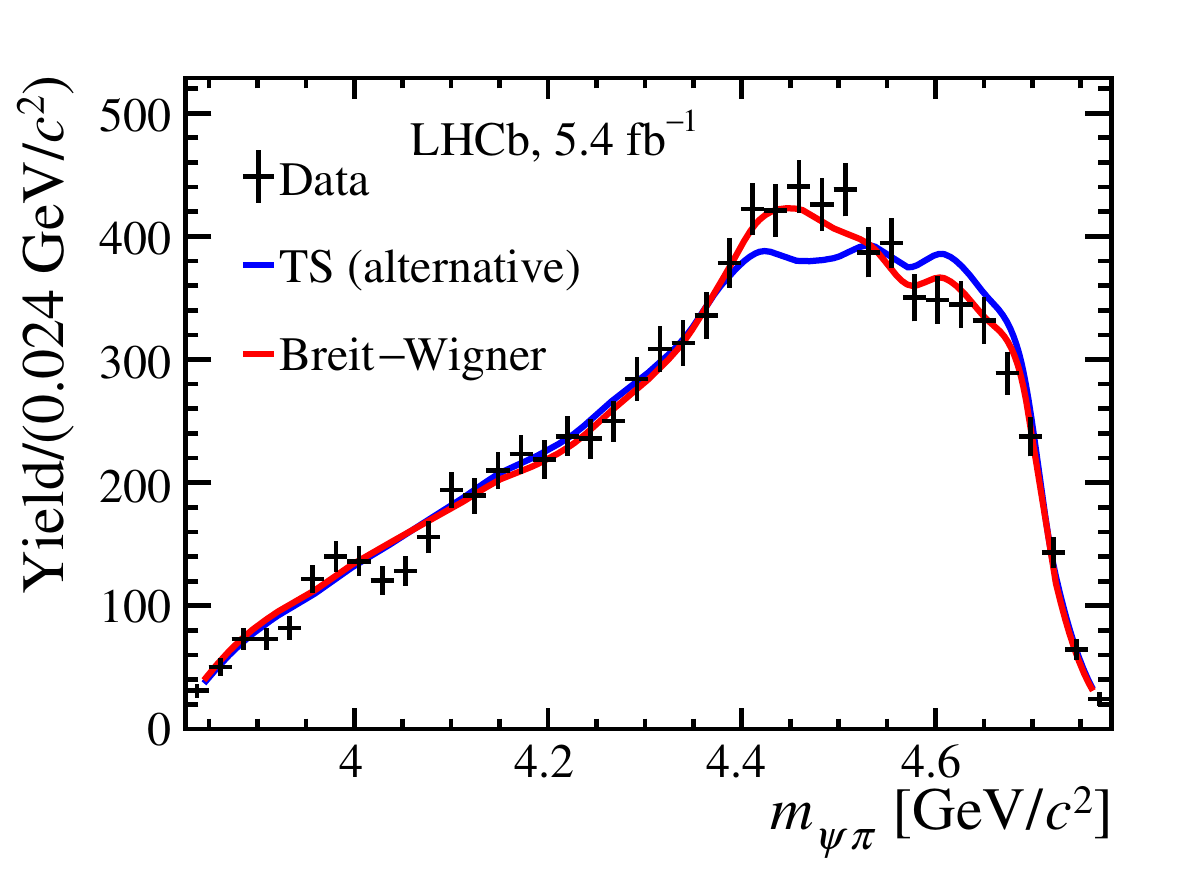}

\end{center}
\caption{
Distribution of $\psi(2S)\pi$ mass of background-subtracted $B^{+} \to \psi(2S) K_{\text{S}}^{0} \pi^{+}$ candidate decays with the result of the amplitude fit for (red) the relativistic Breit--Wigner parametrization and  (blue) the alternative triangle singularity amplitude also shown.}
\label{fig:alterTS}
\end{figure}

Figure~\ref{fig:alterTS} shows the $\psitwos\pip$ invariant-mass projection of the amplitude fit including the alternative triangle singularity amplitude, described in Ref.~\cite{Gribov:2009cfk}.

\clearpage

%% file: Authorship_LHCb-PAPER-2025-039.tex
\centerline
{\large\bf LHCb collaboration}
\begin
{flushleft}
\small
R.~Aaij$^{38}$\lhcborcid{0000-0003-0533-1952},
A.S.W.~Abdelmotteleb$^{57}$\lhcborcid{0000-0001-7905-0542},
C.~Abellan~Beteta$^{51}$\lhcborcid{0009-0009-0869-6798},
F.~Abudin{\'e}n$^{57}$\lhcborcid{0000-0002-6737-3528},
T.~Ackernley$^{61}$\lhcborcid{0000-0002-5951-3498},
A. A. ~Adefisoye$^{69}$\lhcborcid{0000-0003-2448-1550},
B.~Adeva$^{47}$\lhcborcid{0000-0001-9756-3712},
M.~Adinolfi$^{55}$\lhcborcid{0000-0002-1326-1264},
P.~Adlarson$^{85}$\lhcborcid{0000-0001-6280-3851},
C.~Agapopoulou$^{14}$\lhcborcid{0000-0002-2368-0147}
% ...

{\footnotesize \it

% ...
$^{14}$Universit{\'e} Paris-Saclay, CNRS/IN2P3, IJCLab, Orsay, France\\
% ...
$^{38}$Nikhef National Institute for Subatomic Physics, Amsterdam, Netherlands\\
% ...
$^{47}$Instituto Galego de F{\'\i}sica de Altas Enerx{\'\i}as (IGFAE), Universidade de Santiago de Compostela, Santiago de Compostela, Spain\\
% ...
$^{51}$Physik-Institut, Universit{\"a}t Z{\"u}rich, Z{\"u}rich, Switzerland\\
% ...
$^{55}$H.H. Wills Physics Laboratory, University of Bristol, Bristol, United Kingdom\\
% ...
$^{57}$Department of Physics, University of Warwick, Coventry, United Kingdom\\
% ...
$^{60}$School of Physics and Astronomy, University of Glasgow, Glasgow, United Kingdom\\
$^{61}$Oliver Lodge Laboratory, University of Liverpool, Liverpool, United Kingdom\\
% ...
$^{69}$Syracuse University, Syracuse, NY, United States\\
% ...
$^{85}$Department of Physics and Astronomy, Uppsala University, Uppsala, Sweden, associated to $^{60}$\\
% ...
}
\end{flushleft}